\documentclass[useAMS,usenatbib]{mn2e}

\usepackage{graphicx}
\usepackage{times}
\usepackage{natbib}

\newcommand{\apjs}{ApJS}

\newcommand{\apjl}{ApJL}

\newcommand{\aap}{A \& A}

\newcommand{\enzo}{\it{\small ENZO}}

\newcommand{\nvidia}{\it{\small NVIDIA}}
\begin{document}

\title[Magnetic fields in filaments and galaxy clusters] {On the amplification of magnetic fields in cosmic filaments and galaxy clusters{\thanks {F.V. would like to dedicate this paper to the memory of Piero Pasini, who recently passed away, former president of Associazione Astrofili Vittorio Veneto, and old friend and insiprational figure in F.V.'s career.}}}
\author[F. Vazza, M. Br\"{u}ggen, C. Gheller, P. Wang]{F. Vazza$^{1,2}$\thanks{E-mail: franco.vazza@hs.uni-hamburg.de},  M. Br\"{u}ggen$^{1}$, C. Gheller$^{3}$, P. Wang $^{4}$\\
$^{1}$ Hamburger Sternwarte, Gojenbergsweg 112, 21029 Hamburg, Germany\\
$^{2}$INAF/Istituto di Radioastronomia, via Gobetti 101, I-40129 Bologna,
Italy\\
$^{3}$  CSCS, Via Trevano 131, CH-6900 Lugano, Switzerland\\
$^{4}$  {\nvidia}, Santa Clara, 95050, US}

\date{Accepted 2014 September 9. Received 2014 September 8; in original form 2014 September 2.}
\maketitle
\begin{abstract}

The amplification of primordial magnetic fields via a small-scale turbulent dynamo during structure formation might be able to explain the observed magnetic fields in galaxy clusters. The magnetisation of more tenuous large-scale structures such as cosmic filaments is more uncertain, as it is challenging for numerical simulations to achieve the required dynamical range. 
In this work, we present magneto-hydrodynamical cosmological simulations on large uniform grids to study the amplification of primordial seed fields in the intracluster medium (ICM) and in the warm-hot-intergalactic medium (WHIM).
In the ICM, we confirm that turbulence caused by structure formation can produce a significant dynamo amplification, even if the amplification is smaller than what is reported in other papers. In the WHIM inside filaments, we do not observe significant dynamo amplification, even though we achieve Reynolds numbers of $R_{\rm e} \sim 200-300$. The maximal amplification for large filaments is  of the order of $\sim 100$ for the magnetic energy, corresponding to a typical field of a few $\sim \rm nG$ starting from a primordial weak field of $10^{-10}$ G (comoving).  In order to start a small-scale dynamo, we found that a minimum  of $\sim 10^2$ resolution elements across the virial radius of galaxy clusters was necessary. In filaments we could not find a minimum resolution to set off a dynamo. This stems from the inefficiency of supersonic motions in the WHIM in triggering solenoidal modes and small-scale twisting of magnetic field structures. Magnetic fields this small will make it hard to detect filaments in radio observations.

\end{abstract}

\label{firstpage} 


\section{Introduction}
\label{sec:intro}

Cosmic magnetism is an astrophysical puzzle.  While radio observations provide evidence for magnetic field strengths of up to a few $\sim \rm \mu G$ in galaxy clusters and galaxies \citep[e.g.][and references therein]{fe08,br11,ry11}, the origin of such strong fields is unclear, given that the upper limits on the primordial magnetic field at the epoch of the Cosmic Microwave Background set $B<10^{-10}$ G (e.g. \citealt{2010Sci...328...73N}). \\
 From a theoretical point of view, the first cosmic seed fields can be generated in the very early Universe during inflation and first-order phase transitions (however, the uncertainty on the efficiency of such mechanisms is large, $B \sim 10^{-34}-10^{-10}$ G, e.g. \citealt{wi11}). Additional processes such as the Biermann-battery, or aperiodic turbulent fluctuations in the intergalactic plasma, might also provide seed fields in the range $\sim 10^{-19}-10^{-16}$ G \citep[][]{1997ApJ...480..481K,2012PhRvL.109z1101S}.
Later on, structure formation can cause further amplification via a small-scale turbulent dynamo \citep[e.g.][]{su06} in two main phases: first, via exponential growth of the magnetic field in the kinematic regime, and, second, via non-linear growth and stretching of the coherence scales until saturation with the turbulent forcing \citep[][]{wa09,beck12,2013A&A...560A..87S,2014ApJ...783L..20P}. Galactic activity can yield localised additional seeding \citep[e.g.][]{Kronberg..1999ApJ,Volk&Atoyan..ApJ.2000}, while further amplification in cluster outskirts might be produced via the magneto-thermal instability \citep[][]{2008ApJ...688..905P} or instabilities driven by cosmic rays accelerated by shocks \citep[][]{2012MNRAS.427.2308D,brug13}.     
At higher redshifts ($z \sim 2$), star formation should be able to induce small-scale dynamo by injecting turbulence from supernova explosions \citep[e.g.][]{1996ARA&A..34..155B,beck13}, producing large Rotation Measures  \citep[e.g.][]{2008ApJ...676...70K,2012ApJ...755...21M} and possibly explaining the tight correlation between far-infrared and radio continuum emission \citep[][]{2013A&A...556A.142S}.
 \\
Cosmological simulations can reproduce the observed field strengths within galaxies and galaxy clusters starting from weak primordial fields \citep[e.g.][]{do99,br05,bo11,ruszkowski11}, yet similar field strengths can also be achieved with outflows from active galactic nuclei \citep[e.g.][]{xu09,2008A&A...482L..13D}, galactic winds \citep[e.g.][]{donn09} and
 star formation \citep[e.g.][]{beck13}. In particular, \citet{donn09} concluded that magnetized galactic outflows and their subsequent evolution within the ICM in principle can explain the observed magnetisation of galaxy clusters, while measuring cosmological magnetic fields in low-density environments can reveal the origin of cosmic magnetic fields.\\
Very little is known about the evolution and present-day distribution of magnetic fields in the periphery of galaxy clusters and in the cosmic web, particularly in filaments that contain $\sim 50-60$ percent of the total mass in the Universe \citep[e.g.][]{2014MNRAS.441.2923C}.
This circumstance makes the study of ultra-high energy cosmic rays (UHECRs) very uncertain since large-scale magnetic fields change the arrival direction of UHECRs
\citep[e.g.][]{2003PhRvD..68d3002S,2010ApJ...710.1422R}. This also adds uncertainties to the composition of UHECRs, as the presence of magnetic fields can significantly alter the spectrum and composition of UHECRs that reach Earth \citep[e.g.][]{2014NIMPA.742..245A}.\\

Numerical simulations are crucial for studying the non-linear processes that lead to the amplification of the seed magnetic fields during structure formation.
In simulations, a large spatial resolution is needed to produce the degree of turbulence that leads to dynamo amplification  \citep[e.g.][]{2011PhRvL.107k4504F,2012ApJ...745..154T,2013MNRAS.432..668L}. However, in filaments, neither Lagrangian  (such as smooth particle hydrodynamics) nor mesh refinement schemes based on matter density, achieve the necessary resolution.  On the other hand, the use of fixed grids is computationally demanding due to the need of 
resolving the details of the internal structure of filaments. Whether the magnetic field in filaments approaches equipartition with the kinetic energy, is unclear. The amplification is expected to depend on the
numerical resolution, on the exact distribution of modes (compressive or solenoidal), as well as on the range of dynamical scales \citep[][]{2004ApJ...612..276S,ry08,2009ApJ...693.1449C,2011arXiv1108.1369J}.\\

Modelling of magnetic fields in filaments is relevant for the study of radio emission 
from the cosmic web, that surveys in the nearby (e.g. LOFAR) and more distant future (e.g. the SKA) might be able to detect for the
first time, in case of large enough magnetic fields \citep[][]{2011JApA...32..577B,2012MNRAS.423.2325A}. 

\section{Methods}
\label{sec:methods}

\subsection{{\enzo}-MHD}
The simulations performed in this work have been produced with 
a customised version of the grid code {\enzo} \citep[][]{enzo13}.
{\enzo} is a highly parallel code for cosmological magneto-hydrodynamics (MHD), which uses a particle-mesh N-body method (PM) to follow the dynamics of the DM and a variety of shock-capturing Riemann solvers  to evolve the gas component. 

The MHD implementation of {\enzo} that we use has been developed by \citet{wa09} and \cite{wang10}. It is based on the Dedner formulation of MHD equations \citep[][]{ded02}, which uses hyperbolic divergence cleaning to preserve the $\nabla \cdot {\bf B=0}$ condition. The MHD solver adopted here uses a piecewise-linear reconstruction, where fluxes at cell interfaces are calculated using the Harten-Lax-van Leer (HLL) approximate Rimeann solver \citep[][]{1983JCoPh..49..357H} and time integration is performed using a total variation diminishing (TVD) second order Runge-Kutta (RK) scheme \citep[][]{1988JCoPh..77..439S}. The resulting solver is expected to be slightly more diffusive than the piecewise-parabolic approach, but allows a more efficient treatment of the electromagnetic terms. Extensive tests have been conducted to compare the performance of different MHD solvers in astrophysical codes (including the implementation of the Dedner scheme employed here) in the case of decaying supersonic turbulence.  Overall, the Dedner cleaning compared well with more complex MHD schemes, at the price of being more dissipative at very small spatial scales, due to the small-scale $\nabla \cdot {\bf B}$ waves generated by this scheme \citep[][]{kr11}.  For further tests on the validation of the code we refer the reader to \citet[][]{wa09}.

 This MHD solver, as well as a version of the piecewise parabolic method (PPM) hydro solver, has
been ported to {\nvidia}'s CUDA framework, allowing {\enzo} to take advantage of modern graphics hardware \citep[][]{wang10,enzo13}. A key step in {\enzo}'s implementation is flux correction, which is required
when each level of resolution is allowed to take its own time step. Within the GPU version of the MHD solvers, the fluxes are calculated on the GPU and only the fluxes required for flux correction are transferred back to the CPU. This procedure reduces the overhead associated with the data transfer, which can be large in a heterogeneous architecture of this sort. Due to the explicit, directionally-split stencil pattern of both the PPM and Dedner MHD solvers, they are well-suited for hardware acceleration. The porting onto GPUs replaced many shared temporary arrays of the CPU version into larger
temporary arrays that are not shared among loop iterations, and exposed the massive parallelism in the algorithm using CUDA. For further details on the porting onto CUDA, we refer the reader to \citet[][and]{wang10,enzo13}.\\

Most of our simulations were run on the Piz Daint system (deployed by ETHZ CSCS Swiss national supercomputing centre in Lugano \footnote{http://www.cscs.ch/computers/piz\_daint/index.html}, a Cray XC30 supercomputer accounting for more than 5000 computing nodes, each equipped with an 8-core 64-bit Intel SandyBridge CPU (Intel Xeon E5-2670) and an {\nvidia} Tesla K20X GPU. 
When running at fixed mesh resolution, the GPU allow to gain a factor of $\sim 4$ in performance, compared to the usage of the corresponding CPU, reducing accordingly the necessary computing time and allowing the investigation of a larger parameters space, with a given amount of computational resources. 

In the Appendix we present a number of tests performed using the CUDA implementation of {\enzo}'s MHD solver, where we simulated the amplification of a weak uniform field in a cubic box with a steady driving of turbulence.

\begin{figure*}
\begin{center}
\includegraphics[width=0.99\textwidth]{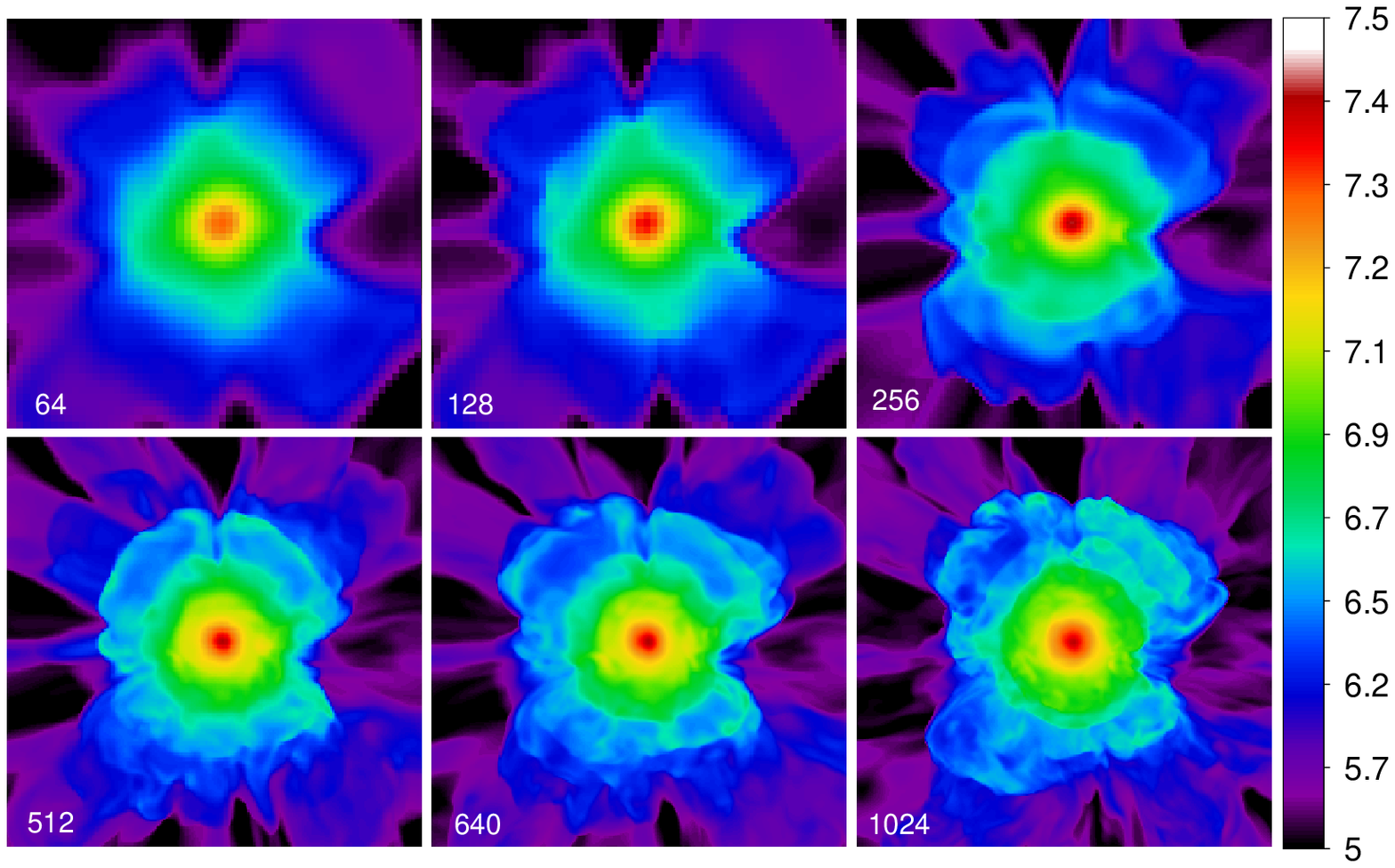}
\includegraphics[width=0.99\textwidth]{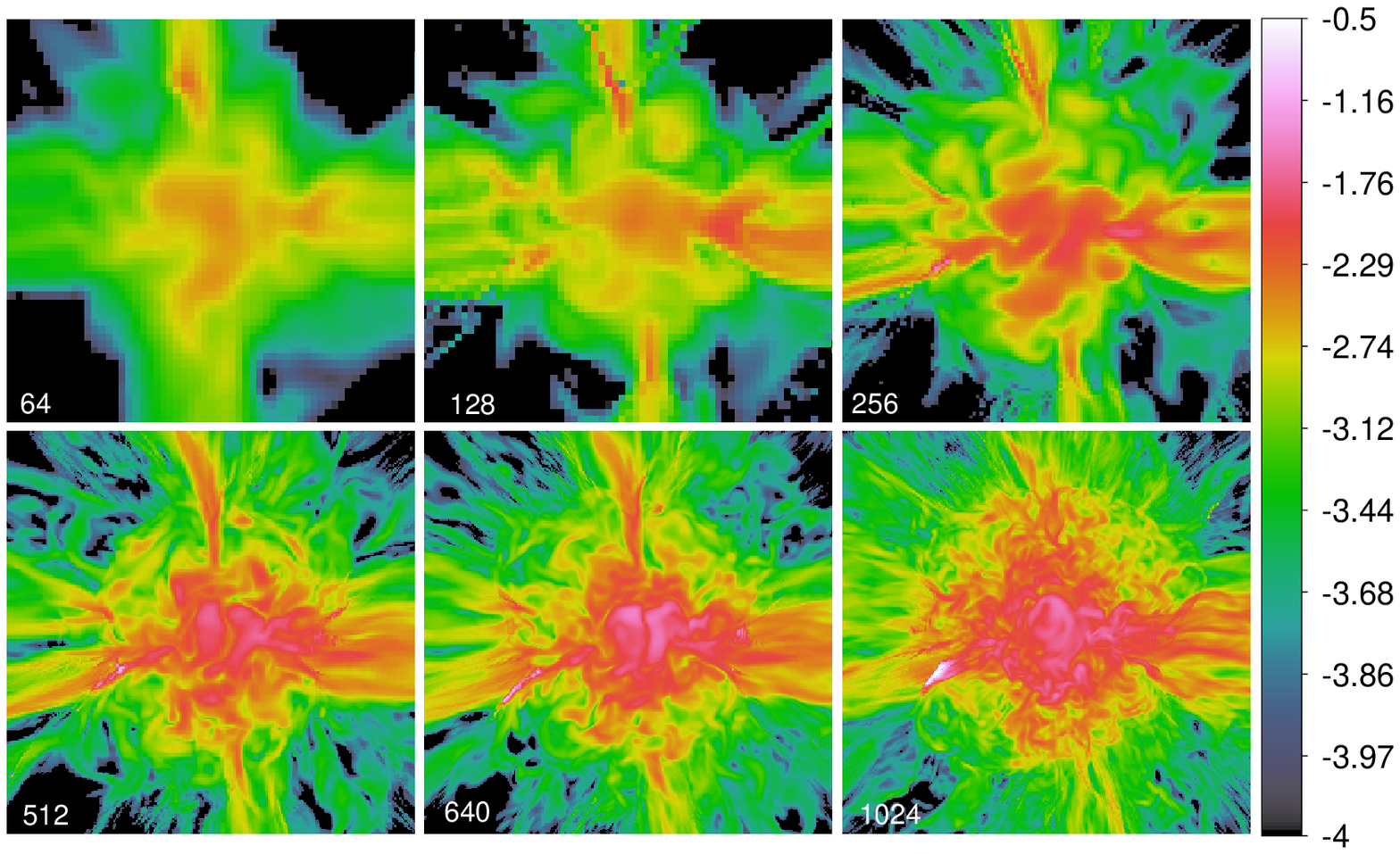}
\caption{Slices through the centre of small (12  Mpc)$^3$ set of fixed grid runs centred on a $\sim 10^{14} M_{\odot}$ cluster, showing the $log_{10}(T[\rm K])$ for increasing resolutions (top) and the average magnetic field strength along the line of sight ($\log_{10}B[\mu G]$, bottom).}
\label{fig:cluster_cut}
\end{center}
\end{figure*}

\subsection{Setups}

We assume a WMAP 7-year cosmology with $\Omega_0 = 1.0$, $\Omega_{B} = 0.0455$, $\Omega_{DM} =
0.2265$, $\Omega_{\Lambda} = 0.728$, Hubble parameter $h = 0.702$,  and a spectral index of $n_s=0.961$ for the primordial spectrum of initial matter
fluctuations \citep[][]{2011ApJS..192...18K}.
The amplitude of the variance of the cosmic spectrum of density at the start of each run has been varied from run to run as explained in Sec.~
\ref{subsec:cluster}-\ref{subsec:cosmo}.
The magnetic field in all runs has been initialised to the reference value of $B_0=10^{-10}$ G (comoving), which we imposed as a background uniform field at the beginning of each run. A list of runs is given in Tab.~\ref{tab:tab1}.

\begin{table}
\label{tab:tab1}
\caption{Main parameters of our MHD runs of a galaxy cluster (with exaggerated $\sigma_{\rm 8}$), employed in Sec.~\ref{subsec:cluster}. 
 First column: number of grid(s) cells in the initial conditions;  column 2: spatial resolution; column 3: dark matter mass resolution.}
\centering \tabcolsep 5pt 
\begin{tabular}{c|c|c}
  $N_{\rm grid}$ & $\Delta x$[kpc] &  $m_{\rm DM} [M_{\odot}/h]$ \\  \hline 
 $64^3$ & $220$ & $2.9 \cdot 10^{8} $ \\
 $128^3$ & $110$ & $3.6 \cdot 10^{7} $ \\
 $256^3$ & $55$ & $4.5 \cdot 10^{6} $ \\ 
 $512^3$ & $27$ & $5.6 \cdot 10^{5} $ \\ 
 $640^3$ & $22$ & $2.9 \cdot 10^{5} $ \\
 $1024^3$ & $13$ & $7.0 \cdot 10^{4} $ \\
 \end{tabular}
\end{table}

\section{Results}
\label{sec:res}

\subsection{Magnetic field amplification in the ICM}
\label{subsec:cluster}

In a first set of simulations, we measured the amplification of a cosmological weak magnetic field during the formation of a galaxy cluster, as a benchmark test for our following studies of amplification within filaments with {\enzo}-MHD. \\
This magnetisation of the ICM during structure formation has already been studied with a variety of  codes by many authors \citep[e.g.][]{do99,br05,2008A&A...482L..13D,xu09,co11,bo11}, that demonstrated how the amplification of magnetic field is a natural process within the
large over-density of galaxy clusters (even if the amplification factors can change from simulation to simulation). 
Here we want to study the growth of magnetic fields as a function of spatial resolution. Hence, we want to limit as much as possible the uncertainties related to the use of adaptive mesh refinement \citep[e.g.][]{xu09}. Therefore, we only used runs with uniform spatial resolution along the whole cluster evolution.To this end, we adopted an
artificially large normalisation of the primordial matter power spectrum, $\sigma_{8}=5.0$, in creating our initial conditions, in order to enable the formation of a single
cluster of mass $\sim 10^{14} M_{\odot}$ even within the rather small volume of (14  Mpc)$^3$. 
Of course, this unrealistically large value of $\sigma_{8}$ (to be compared with the concordance value $\sigma_{8} \approx 0.8$) will shed little light on the timing of the amplification since a large value of $\sigma_{8}$ 
causes the formation of clusters already at high redshifts.
Using this idealised setup, we simulated the evolution of the ICM employing grids from 
$64^3$ to $1024^3$ cells/DM particles corresponding to a comoving spatial resolution from  
$220$ kpc to $13$ kpc. A list of our cluster runs is given in Tab.~\ref{tab:tab1}.  
 
Figure \ref{fig:cluster_cut} shows maps of temperature and magnetic fields for a slice through the centre of the cluster at $z=0$, for all resolutions from $64^3$ to $1024^3$.
While the temperature distribution of the cluster varies slightly across runs, the spatial distribution
of the magnetic fields changes clearly with increasing resolution. Starting at a resolution
of $27$ kpc ($512^3$) the morphology of the magnetic field becomes increasingly more tangled on scales smaller than the cluster core radius, and clumps of gas with $B \geq 0.1~ \mu G$ start to appear throughout the virial volume. At our best resolution, the maximum Reynolds number within the virial volume is:

\begin{equation}
 R_{e} \approx  (\frac{2 R_{\rm v}}{2 \Delta x})^{4/3} \approx 1400
\label{eq:reynolds}
\end{equation}
where $R_{\rm v}=1.5$ Mpc is the virial radius at $z=0$ \citep[e.g.][]{va11turbo} and $\Delta x$ is our (comoving) spatial resolution ($13$ kpc in the most resolved run). According to simulations of forced turbulence in a box \citep[][]{2004ApJ...612..276S,2009ApJ...693.1449C,2011arXiv1108.1369J}, this is large enough to start a small-scale dynamo. 
The former likely represents an overestimate of the real Reynolds number in the flow, because the cluster's virial radius was smaller in the past, and because the driving of the turbulence by sub-clusters preferentially occurs on scales smaller than the current virial radius \citep[][]{va12filter}, thereby limiting the outer scale of turbulence.\\    

\begin{figure}
\begin{center}
\includegraphics[width=0.49\textwidth]{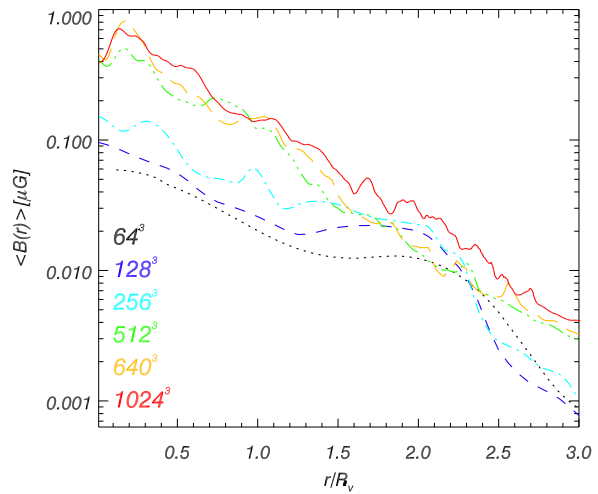}
\caption{Average profile of the magnetic field at $z=0$ for our cluster runs at all resolutions.}
\label{fig:profile}
\end{center}
\end{figure}

At all resolutions, the radial profile of the magnetic fields at $z=0$ (Fig.~\ref{fig:profile}) shows the build-up of the 
magnetic field in the centre. The growth of the field proceeds faster with increasing resolution in the innermost regions. 															
Inside the virial volume, the average profile of the magnetic field does not vary much with resolution in the range between 27 kpc and 13 kpc, suggesting that we are not far from convergence.  The maximum field we observe in the centre is $\sim 0.7 \mu G$, corresponding to  a maximum amplification factor of $\sim 5 \cdot  10^7$ for the magnetic energy and 7000 for the magnetic field. 
Beyond the virial radius the simulation does not seem to be fully converged. At distances of 1 $R_{\rm v}$ from the cluster centre, the average field varies from $\sim 0.02-0.04~ \mu$G at low resolution to $\sim 0.1 \mu$G at high resolution. 
For a cluster of this mass and central temperature ($\sim 3 \cdot 10^{7} \rm K$), the resulting plasma beta is
of the order of $\beta \sim 100$ (where $\beta=n k_{B} T /P_{\rm B}$, where $n$ is the gas density and $P_{\rm B}$ is the magnetic pressure) in the innermost cluster regions. This matches observations for real galaxy clusters \citep[][]{mu04,bo10}.\\

\begin{figure}
\begin{center}
\includegraphics[width=0.49\textwidth]{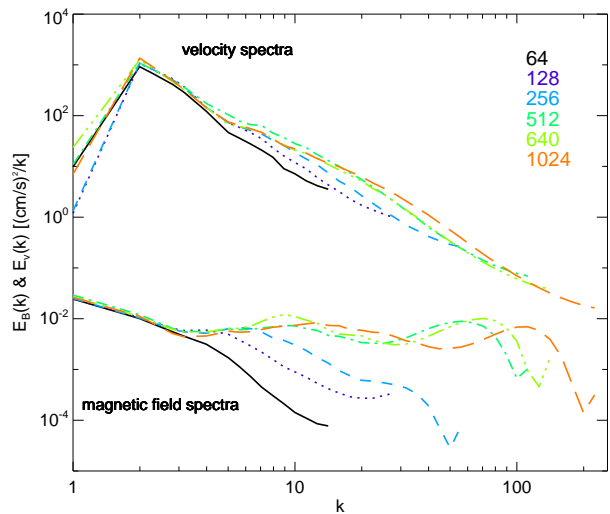}
\caption{Specific kinetic energy (top lines) and magnetic (lower lines) spectra
for a volume of (7 Mpc)$^3$ centre of the cluster of Fig.~\ref{fig:cluster_cut} for all simulated resolutions. The spatial frequency, $k$, is in units of the box size and for each run goes from $k=1$ ($7  ~\rm Mpc$) to the Nyquist frequency of each spectrum (i.e. twice the grid resolution of each run).}
\label{fig:spectra_cluster_res}
\end{center}
\end{figure}

\begin{figure}
\begin{center}
\includegraphics[width=0.49\textwidth]{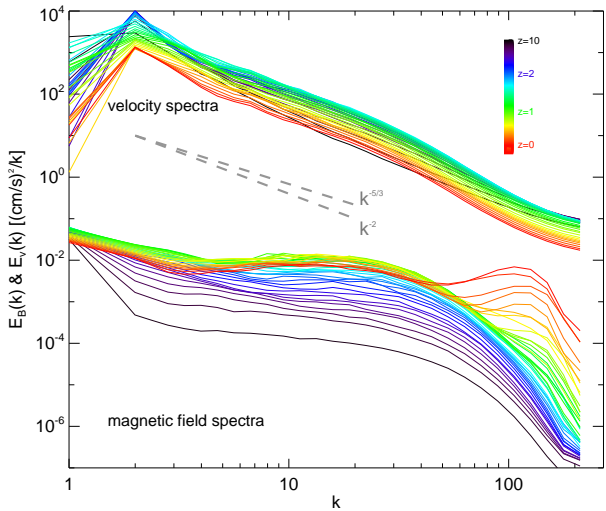}
\caption{Evolution of  (comoving) velocity (top lines) and magnetic (lower lines) spectra
for the $1024^3$ cluster run. The additional horizontal lines show the reference slopes of $k^{-5/3}$ and $k^{-2}$.}
\label{fig:spectra_cluster_1024}
\end{center}
\end{figure}

Figure \ref{fig:spectra_cluster_res} shows the comoving kinetic energy per unit mass (top lines) and magnetic
field spectra (lower lines) for all resolutions at $z=0$. All spectra were computed in a (7 Mpc)$^3$ cubic box centred on the cluster, using an FFT algorithm and assuming periodic boundary conditions. In order to compare our spectra to standard ``turbulence in a box'' simulations \citep[e.g.][]{2003ApJ...597L.141H,2004ApJ...612..276S,2009ApJ...693.1449C,kr11}, we assumed $\rho=1$ for the gas, which removes the effect of density fluctuations on the kinetic energy spectra.
The specific kinetic energy spectra are very similar at all resolutions, with a power-law
slightly steeper than the Kolmogorov slope across more than two orders of magnitude in scale. This is in agreement with previous numerical results \citep[e.g.][]{va09turbo,va11turbo,va12filter,gaspari13}.
The magnetic field spectra, however, show the clear build-up of the small-scale magnetic field as soon as the spatial resolution is sufficiently fine.
From $k \geq 4$ the magnetic spectra get shallower as resolution
is increased, and in the range $10 \leq k \leq 100$ a significant pile-up of magnetic energy occurs for
resolutions better than $256^3$ (i.e. $\Delta x \leq 55$ kpc). The observed small-scale spectra are qualitatively similar to previous results
by \citet{xu09}, even if their seeding model for the magnetic field differs from that we adopted. No developed power-law spectra is observed for the magnetic field, but a peak that moves towards larger scales as resolution is increased, similar to \citet[][]{2003ApJ...597L.141H,2009ApJ...693.1449C} and at odds with what is usually assumed in Faraday Rotation models \citep[][]{mu04,bo10,bo13}. The peak in the magnetic energy is located at $k \sim 100$ ($\sim 50$ kpc) in our highest resolution run.\\
The build-up over time of the small-scale magnetic field is shown in Figure \ref{fig:spectra_cluster_1024} for our $1024^3$ run.\footnote{We observe that at scales close to the resolution of the box, i.e. $k \sim 256$ in Fig.~\ref{fig:spectra_cluster_1024} a spurious effect on the magnetic spectra is caused by the small-scale waves used by the Dedner cleaning to preserve $\nabla \cdot {\bf B=0}$. We found this effect particularly in the small-scale fields in the cold hypersonic flows outside of clusters and filaments. Here the energy equation is evolved adopting the dual energy formalism \citep[][]{enzo13}, a regime where the exact conservation of hydro/MHD quantities is non-trivial due to the large unbalance between the kinetic and the internal gas energy of cells.} 
The dependence on resolution is stronger for the magnetic field than for the velocity field, and the highest resolution run shows a final magnetic field energy which is a factor $\sim 10^3$ larger than that of the lowest resolution run. 
The small change of the final magnetic energy going from $640^3$ to $1024^3$ (where actually the total magnetic energy is slightly lower, an effect we ascribe to tiny variations in the non-linear evolution of the MHD structure within the volume) suggests
that no further increase in the spatial resolution can produce a significant increase
in the magnetic field amplification.\\

\begin{figure}
\begin{center}
\includegraphics[width=0.49\textwidth]{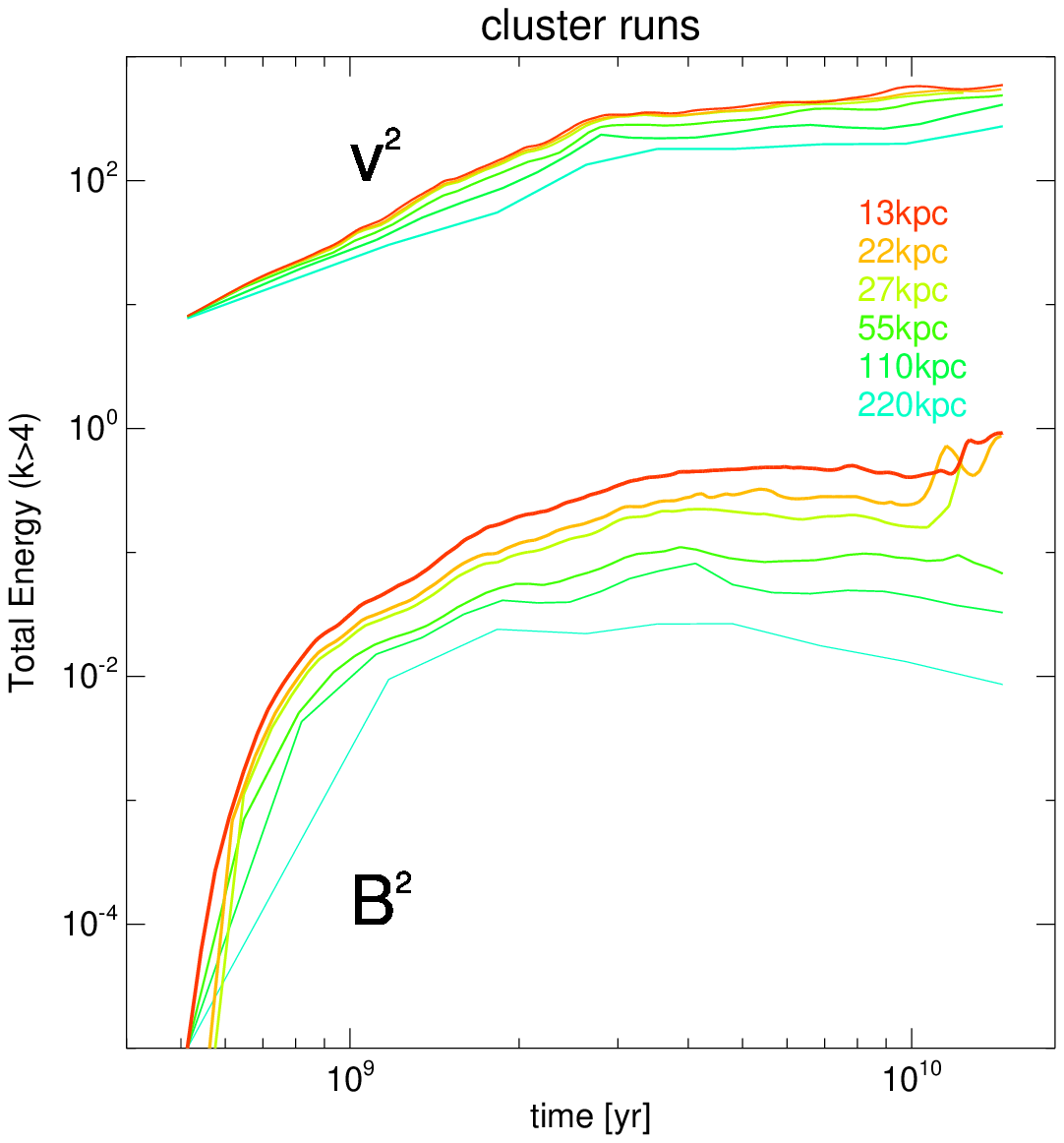}
\caption{Evolution of the total kinetic energy per unit mass inside the cluster volume of Fig.~\ref{fig:cluster_cut} (top lines) and of the total magnetic energy for the same volume, as a function of resolution. The energies are given in [$(cm/s)^2$].}
\label{fig:clust_time_evol}
\end{center}
\end{figure}

Figure \ref{fig:clust_time_evol} shows the evolution of $\int E_{\rm v}(k)dk$ and $\int E_{\rm B}(k) dk$
for all runs, where we integrated the spectra only from $k_{\rm cl} \geq 4$ in order to focus on the
kinetic/magnetic energy fluctuations contained within the cluster volume ($1/k_{\rm cl} \propto R_{\rm v}$, where $R_{\rm v} \sim 1.5$ Mpc at $z=0$).
The total comoving kinetic energy per unit of mass is smaller by one order of magnitude
going from $z=10$ to $z=0$. This is an effect of the thermal dissipation of
infall motions via shock heating and turbulent dissipation, and the 
increase of the small-scale kinetic energy as a function of resolution is only modest,  i.e. a factor $\sim 3$ by $z=0$. On the other hand, the small-scale magnetic energy
is increased by a factor $\sim 10^6$ by the end of the run. Even in this case, the amplified field is far from equipartition with the velocity field
at all scales, even if the difference at the smallest scale is small ($E_{\rm B}/E_{\rm v} \sim 0.1-0.3$ for $k \sim 100$), and in the fully saturated stage the peak of the small-scale magnetic energy is
expected to drift ot even smaller spatial scales \citep[][]{2013MNRAS.429.2469B}. \\

In summary, our tests confirm the start of small-scale turbulent amplification of magnetic fields at high resolution. The typical magnetic field strength reaches a maximum of $\sim 0.7  ~\mu$G in the cluster centre. 
Even if the exact level of the amplification might depend on numerical details and codes
\citep[e.g.][]{do99,br05,xu09,bo10,co11}, our results are in agreement with the basic scenario of turbulent amplification of primordial
fields to explain the observed magnetisation of galaxy clusters.

\begin{table}
\label{tab:tab2}
\caption{Main parameters of our MHD runs of a cosmic filament, referred to in Sec.~\ref{subsec:fila}.  First column: number of grid(s) cells in the initial conditions;  column 2: spatial resolution; column 3: dark matter mass resolution; column 4: AMR levels (D=refinement based on the local gas over density; V=refinement based on the local velocity jump).}
\centering \tabcolsep 5pt 
\begin{tabular}{c|c|c|c|c}
  $N_{\rm grid}$ & $\Delta x$[kpc] &  $m_{\rm DM} [M_{\odot}/h]$ & AMR\\  \hline 
 $64^3$ & $1170$ & $1.22 \cdot 10^{11} $ & 0\\
 $128^3$ & $585$ & $1.52 \cdot 10^{10} $ & 0\\
 $256^3$ & $292$ & $1.90 \cdot 10^{9} $ & 0\\ 
 $512^3$ & $146$ & $2.38 \cdot 10^{8} $ & 0\\ 
 $512^3$ & $73$ & $2.38 \cdot 10^{8} $ & 1D\\ 
 $512^3$ & $73$ & $2.38 \cdot 10^{8} $  & 1DV\\ 
 $512^3$ & $36$ & $2.38 \cdot 10^{8} $  & 2D \\ 
 \end{tabular}
\end{table}

\begin{figure}
\begin{center}
\includegraphics[width=0.485\textwidth]{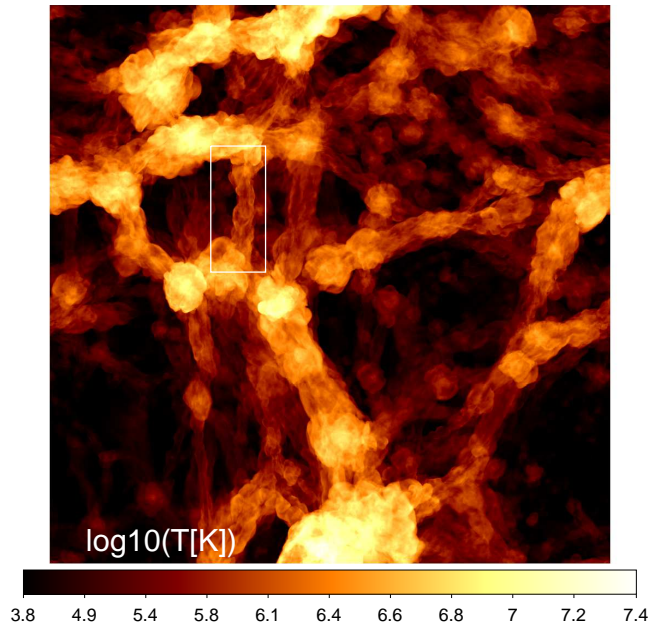}
\caption{Projected (volume-weighted) temperature for the parent $512^3$ simulation of a (75 Mpc)$^3$ volume at $z=0$ that we used to re-simulate the evolution of a massive filament (within the white selection) in Sec.~\ref{subsec:fila}.}
\label{fig:fila_box}
\end{center}
\end{figure}

\begin{figure*}
\begin{center}
\includegraphics[width=0.99\textwidth]{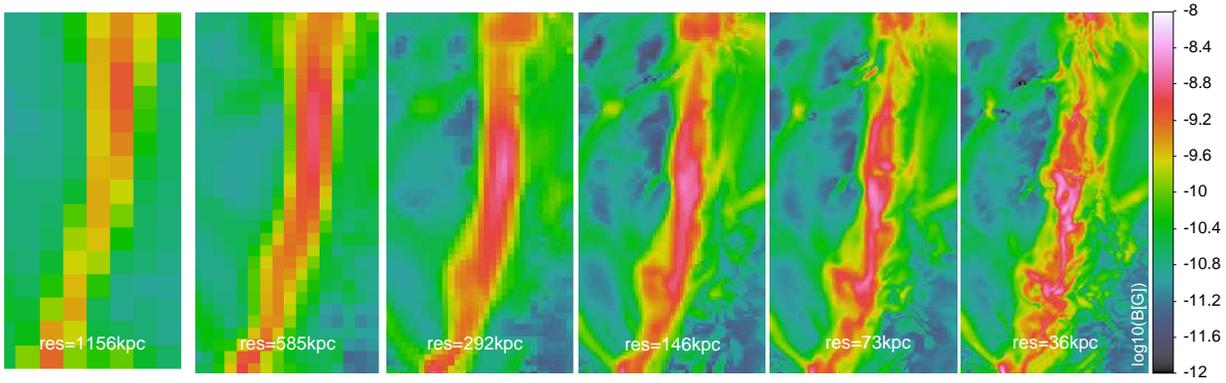}
\caption{Slices of magnetic field strength in a filament with increasing spatial resolution, at $z=0$. The size of each slice is 9 Mpc $\times 18$ Mpc.}
\label{fig:fila_resolution}
\end{center}
\end{figure*}

\begin{figure*}
\begin{center}
\includegraphics[width=0.95\textwidth]{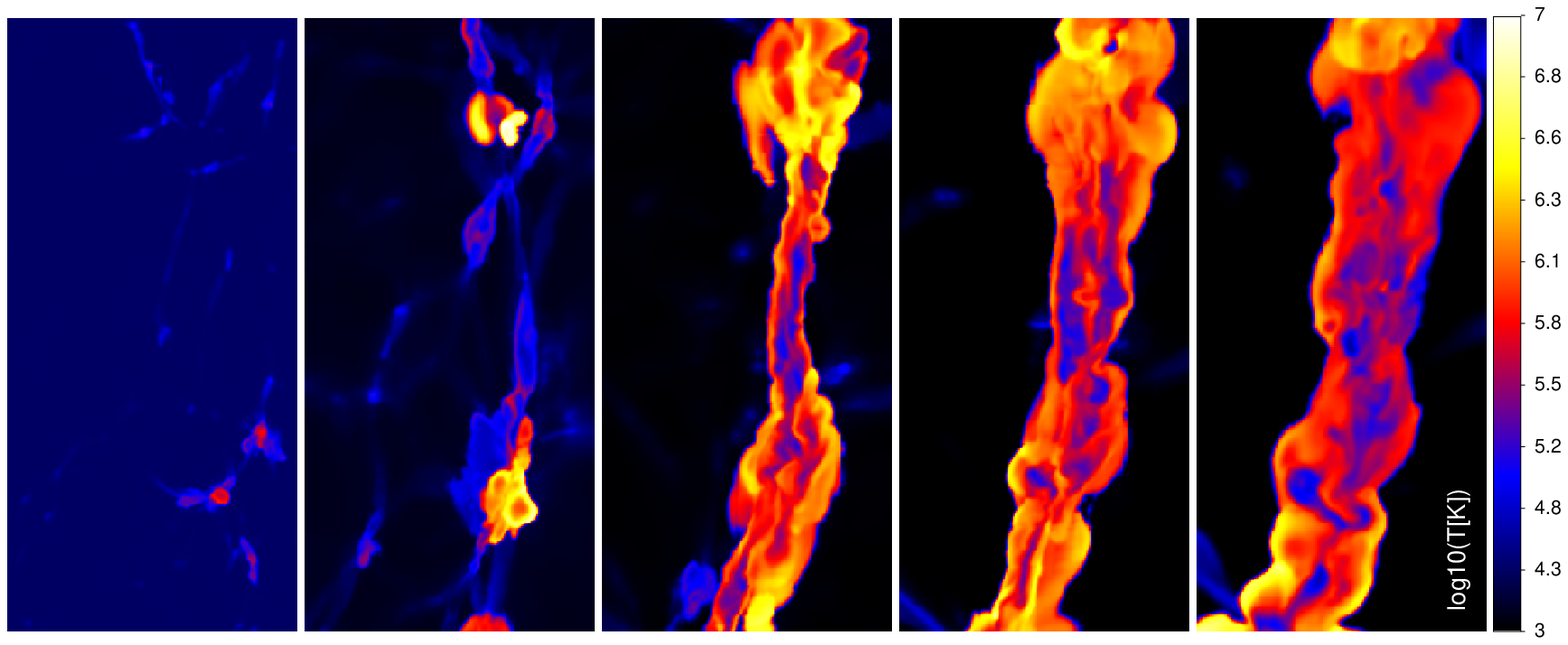}
\includegraphics[width=0.95\textwidth]{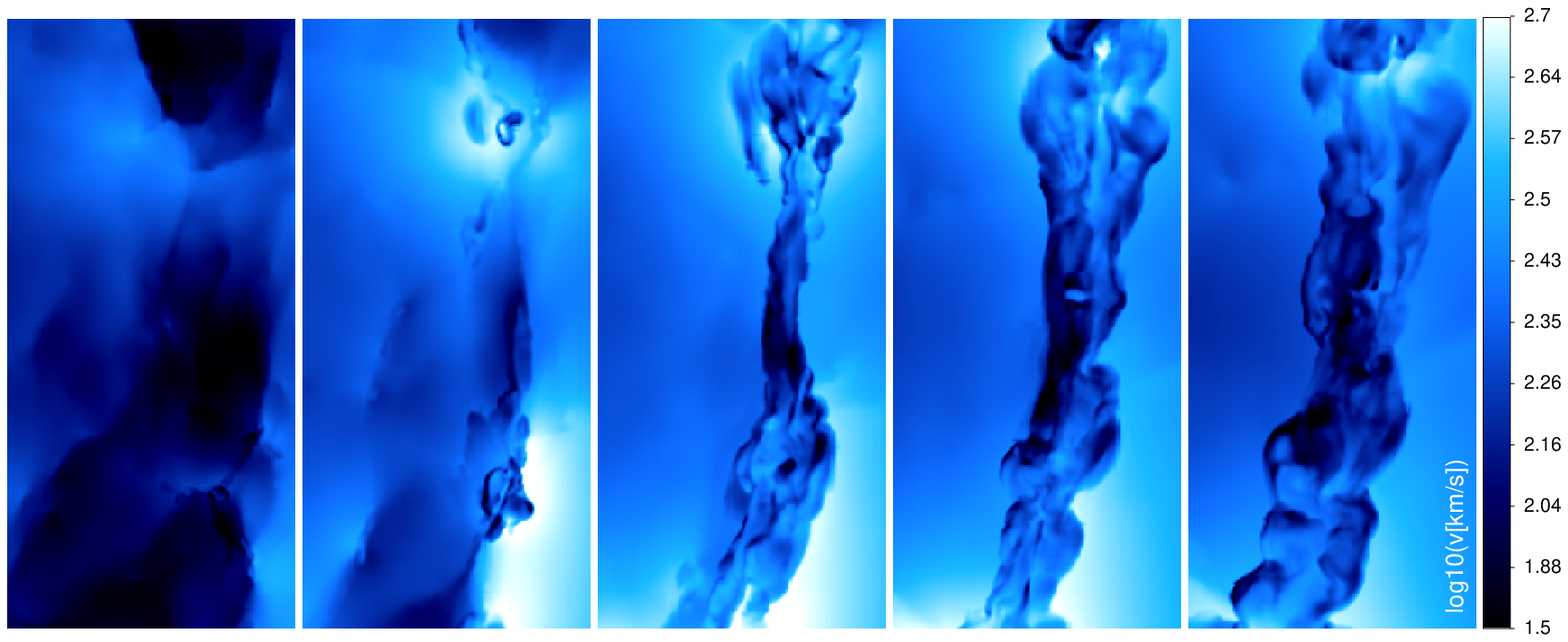}
\includegraphics[width=0.95\textwidth]{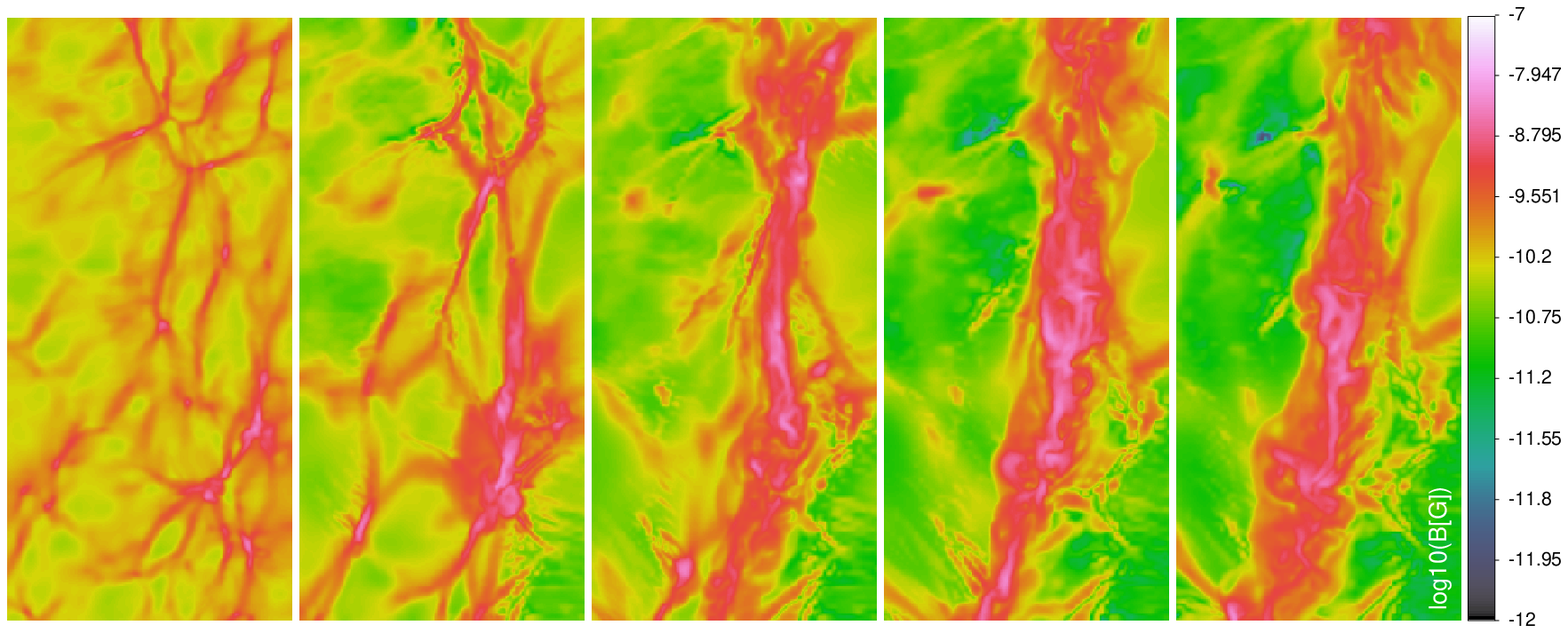}
\caption{Evolution of the filament in a simulation with $512^3$ cells and 2 levels of AMR (the peak resolution is $36$ kpc). From top to bottom, the fields shown are: gas temperature, velocity modulus and magnetic field strength, for a thin slice of depth 36 kpc. The sides of each image measure
9 Mpc $\times 18$ Mpc. From left to right, the redshifts are $z=4.6$, $z=1.8$, $z=0.9$ and $z=0.4$ and $z=0$.}
\label{fig:fila_cut}
\end{center}
\end{figure*}

\begin{figure*}
\begin{center}
\includegraphics[width=0.95\textwidth]{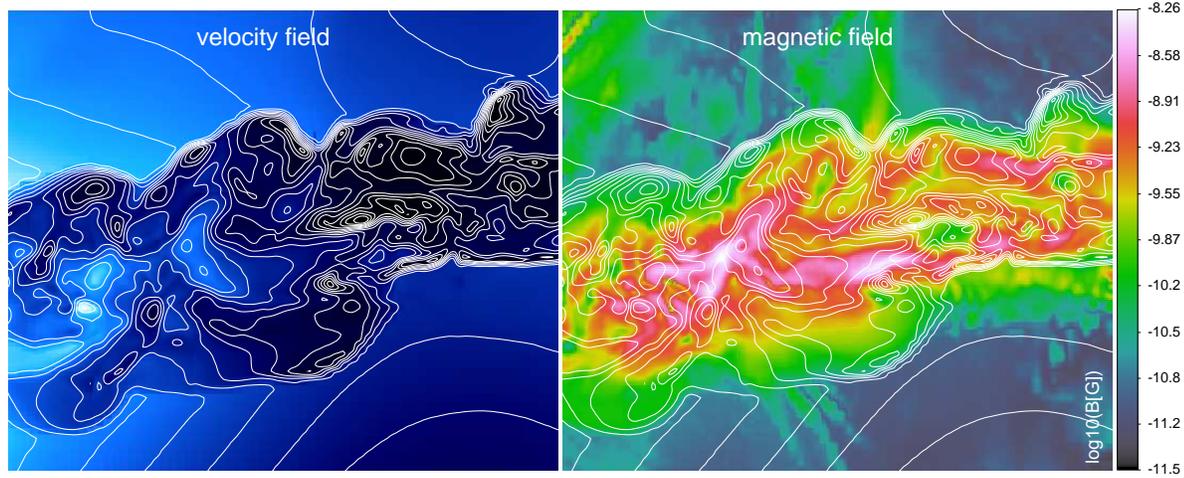}
\caption{Details of the gas velocity flow (left) and of the magnetic field strength (right) for a central slice through the simulated filament of Fig.~\ref{fig:fila_cut}.
The contours are the same in both panels and show the isocontours of the velocity field. The size of the image is $9$ Mpc $\times 7$ Mpc and the peak resolution is $36$ kpc.}
\label{fig:fila_zoom}
\end{center}
\end{figure*}

\begin{figure}
\begin{center}
\includegraphics[width=0.49\textwidth]{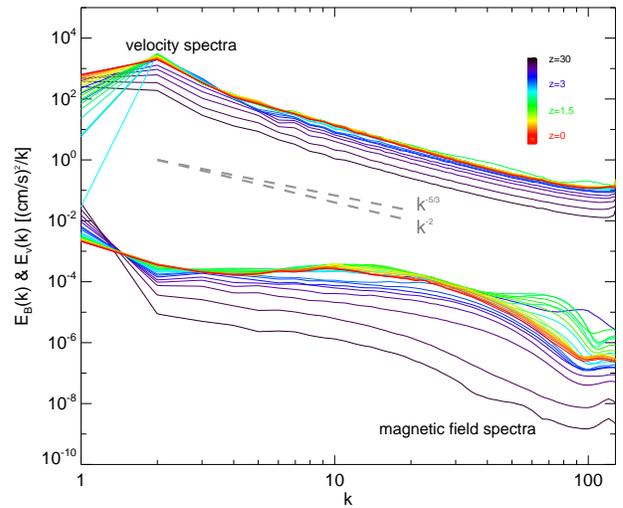}
\caption{Evolution of the comoving velocity spectra (top lines) and of the comoving magnetic spectra (lower lines) for our filament run with $512^3$ cells and 2 AMR levels. The additional horizontal lines show the reference slopes of $k^{-5/3}$ and $k^{-2}$. The spatial frequency, $k$, is in unit of the box size and for each run goes from $k=1$ ($18 ~\rm Mpc$) to the Nyquist frequency of each spectrum (i.e. twice the grid resolution of each run).}
\label{fig:fila_spec_evol}
\end{center}
\end{figure}

\begin{figure}
\begin{center}
\includegraphics[width=0.49\textwidth]{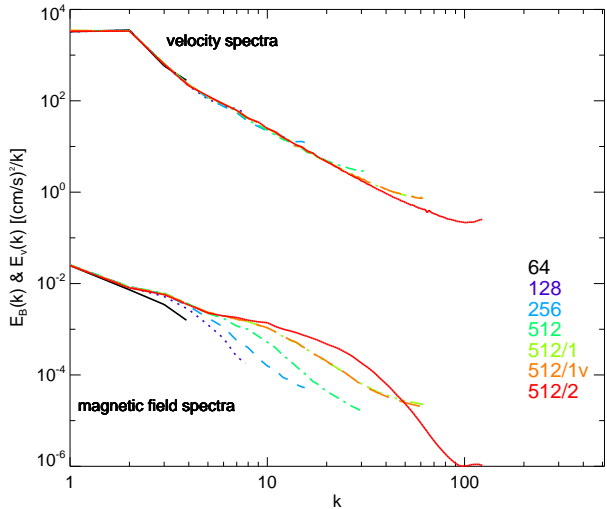}
\caption{Comparison of the velocity spectra (top lines) and of the magnetic spectra (lower lines) for our filament runs (Sec.~\ref{subsec:fila}) at $z=0$ for different resolutions. The units of the spatial frequency $k$ are as in Fig.\ref{fig:fila_spec_evol}.}
\label{fig:fila_spec_res}
\end{center}
\end{figure}

\begin{figure}
\begin{center}
\includegraphics[width=0.49\textwidth]{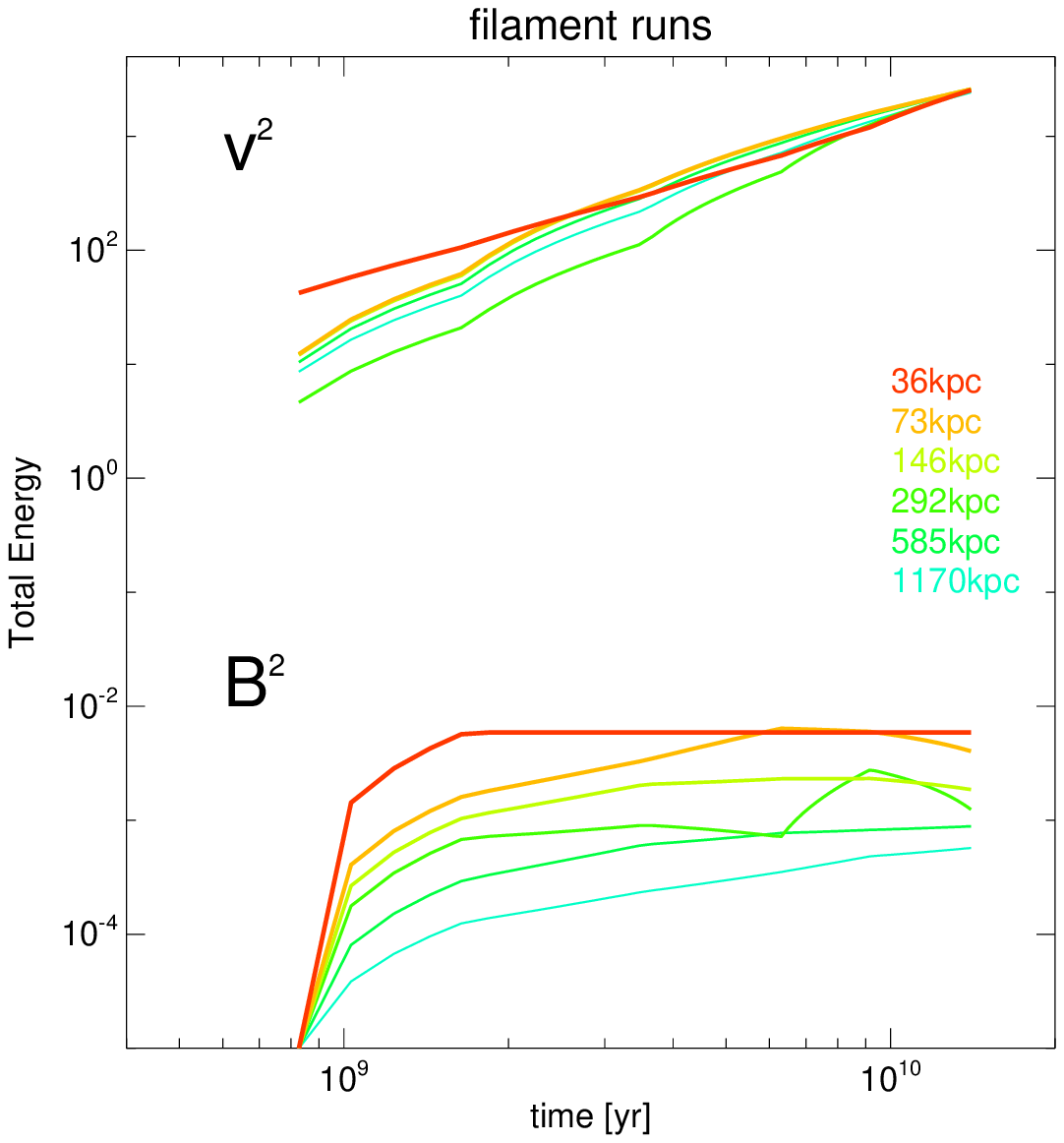}
\caption{Evolution of the total kinetic energy per unit of mass inside the filament volume (top lines) and of the total magnetic energy for the same volume, as a function of resolution. The energies are given in [(cm/s)$^2$]}
\label{fig:fila_time_evol}
\end{center}
\end{figure}

\subsection{Magnetic field amplification in filaments}
\label{subsec:fila}

In a separate set of MHD runs, we investigated the amplification of the primordial magnetic field in a cosmic filament. 
Here we used the canonic value of $\sigma_{\rm 8}=0.8$ and started from a larger cosmological volume (75 Mpc)$^3$,
in which we selected a massive $\sim 15$ Mpc long filament which may be regarded as representative (as shown in the large-scale view of Fig.~\ref{fig:fila_box}). As before, we initialised the primordial field at $z=30$ as a uniform
field with strength $B_0=10^{-10}$ G (comoving). Figure \ref{fig:fila_resolution} shows the final magnetic field in a central slice through all our filament runs at $z=0$. The magnetic
field is the highest close to the major axis of the filament, and its maximum observed strength is only of a few $\sim ~\rm nG$. 
Fig.~\ref{fig:fila_cut} shows slices of  gas temperature, velocity and magnetic field through the centre of the filament along its length taken at different epochs. The filament is already in place at $z=1$ and connects two $\sim 10^{14} \rm M_{\odot}$ clusters (that are located outside of the adaptive mesh refinement, AMR, region). Its peripheral 
regions feature strong ($M \sim 10-100$) accretion shocks along its extension, where the accreted smooth gas (that mostly falls into it along the perpendicular of the accretion region) is shock-heated to a few $\sim 10^6$ K. 
Downstream of accretion shocks inside the filament, most of the gas flow is supersonic, as the sound speed at $T \sim 10^6$ K is $c_s \approx 100$ km/s, lower than the measured velocities (which are $\sim 100-300$ km/s). The magnetic field increases 
from $\sim 3 \cdot 10^{-11}$ G to a few $\sim 10^{-9}$ G downstream of the shocks. After this first boost, there is  little further amplification within the filament and even the most magnetised patches hardly reach $\sim 10^{-8}$ G.
We have highlighted some of these patches in Fig.~\ref{fig:fila_zoom}, where we compared the velocity field and the magnetic field strength within a  slice through the filament. Although there is no one-to-one correlation between velocity field and magnetic field, 
the observed trend suggests that further amplification within the filament occurs in the proximity of
shocks or regions where gas flows collide. However, there is little evidence of eddies with strong curling motions. This is quite different from clusters at comparable resolution. If we rescale the number of cells by the width of the filament, our most resolved run here is comparable to our $1024^3$ cluster run
in terms of the maximum Reynolds number in the flow.\\

In order to test convergence, we re-simulated the same initial conditions with four different resolutions on a fixed grid (from $64^3$ to $512^3$ cells/DM particles). The filament we have chosen is roughly oriented along the $z$-axis of the grid (Fig.~\ref{fig:fila_box}), which also enabled us to perform additional AMR runs by restricting the region for the active refinement to a narrow
rectangular selection within the root grid volume. Thus we have re-simulated the region with up to two more levels of refinement (reaching a maximum resolution of $36 ~\rm kpc$). By the end of its evolution, the filament reaches a transverse size up to $\sim 4$ Mpc, corresponding to $\sim 200$ cells in our AMR runs with three levels. In AMR runs, we let {\enzo} refine the cell size by a factor of two wherever the local gas density exceeded 
the density at the level $l$ by factor $\Delta={\rho_{l}}/\rho_{l-1}$, where we have set $\Delta=3$. 
As previously remarked, the use of AMR may not be optimal for the study of magnetic fields since refining on matter over-density alone can artificially suppress turbulence in regions that are relevant for dynamo amplification \citep[e.g.][]{xu09}. For this reason, in a control run with one AMR level, we also enabled AMR wherever the velocity jump along any of the coordinate axes was larger than $\Delta_v=|v_{j+1}-v_{j-1}|/|v_{j}|$, as in \citet[][]{va09turbo}.  The results are very similar to those 
obtained adopting the density refinement criterion only,
since the turbulent velocity field within the filament is mostly supersonic \citep[][]{ry08}, and the density variations within it are large enough for our conservative choice for $\Delta$ to trigger refinements in most of its interior. It turns out that from redshift $z=1$, roughly 20-25 percent of our AMR region is covered by cells at the highest resolution ($\approx 6.55 \cdot 10^6$ cells), corresponding to more than a $\sim 60-70$ percent of the volume occupied by the filament within the AMR region itself.  Incidentally, the same choice would not work for galaxy clusters, where the density varies more gently and the turbulence is subsonic \citep[][]{xu09,va09turbo}. The parameters for this set of simulations are summarised in Tab.~\ref{tab:tab2}.\\

The velocity spectra for our run with the highest resolution displays a similar evolution as in the cluster (Fig.~\ref{fig:fila_spec_evol}), and present a well-defined power law (compatible with $\propto k^{-2}$) for nearly two decades in scale. The magnetic
spectra again do not show a clear power-law behaviour, and show small-scale bumps which evolve with time. 
However, the build-up of the small-scale magnetic structure is much less significant than in the ICM. Moreover, the trend
does not increase over time but reaches it maximum around $z \sim 1$ (green lines), while the small-scale power is $\sim 1-2$ orders
of magnitude smaller at $z=0$. Overall, the magnetic spectra seem to evolve much faster towards their maximum, compared to the case of the ICM, but 
since $z \sim 1$ they do not show significant evolution on most scales.
The maximum in the magnetic field spectra on small scales ($k \geq 80$, corresponding to $\leq 200$ kpc) is matched by an excess of 
velocity power at the same scales. The time corresponds to the epoch in which the filamentary region that connects the two forming clusters
assembles most of his mass, and when shock heating raises the WHIM's temperature to $\geq 10^6$  K (Fig.~\ref{fig:fila_cut}). At this time, gas flows into the filament from opposite sides at large velocities, and compresses the magnetic fields.  Still, the plasma beta is only of the order of $\beta \sim 10^5-10^6$ in the filament.\\

The dependence on resolution of the power spectra (Fig.~\ref{fig:fila_spec_res}) is similar to that of clusters (Fig.~\ref{fig:spectra_cluster_res}), and  runs with higher resolution show the build-up of small-scale magnetic fields, even if less evident than in the ICM. For comparison, at $k=10$ the magnetic power spectra increases
by more than a factor of $\sim 100$ in the ICM run when the resolution is increased by a factor of 8, while this is less than a factor $\sim 10$ in the filament. \\

The integrated velocity and magnetic spectra as a function of time are given in Fig.~\ref{fig:fila_time_evol}. Again, we filtered out scales smaller than the mean
diameter of the filament at $z=0$, to focus only on velocity and magnetic field fluctuations that
are roughly contained within the filament ($\leq 4$ Mpc). The continuous accretion of matter onto
the filament causes the growth of both quantities: during the
whole evolution the specific kinetic energy has increased by $\sim 2$ orders of magnitude,
while the magnetic energy has increased by $\sim 3$ orders of magnitude in our best resolved runs, and $\sim 2$ orders of magnitude in our coarsest run. This is very different from 
our previous results for clusters (Fig.~\ref{fig:clust_time_evol}). While the increase of specific kinetic 
energy is similar, the increase of magnetic energy with resolution is much slower with resolution, indicating that
convergence might be within reach.  This suggests that starting from a resolution of the order of 73 kpc ($\sim 1/60$ of the thickness of the filament) or better, the effects of compressive modes and shocks on the final magnetic field does not increase with resolution.\\

We conclude that despite the large dynamical range of scales of our AMR runs (corresponding to a Reynolds number of $\sim 210$ in our most resolved case, by assuming that the outer scale is the average diameter of the filament, $\sim 4$ Mpc), in our simulated filament we do not observe a significant small-scale dynamo. Moreover, the trend with resolution of spectra and integrated quantities indicates that the lack of efficient amplification is robust against further increase in resolution, thereby limiting the maximum amplification factor to $\sim 100$ for the magnetic energy in the WHIM for this object. In Sec.~\ref{sec:discussion}, we will discuss this further. 

\subsection{Larger cosmological runs}
\label{subsec:cosmo}

Given the limited statistical significance of results obtained with single objects, we proceed to large-scale unigrid cosmological simulations comprising hundreds of clusters and filaments.  As before, we initialised the magnetic field with a uniform $B_{\rm 0}=10^{10} \rm G$ at $z=30$ and employed the Dedner scheme on the MHD version of {\enzo} \citep[][]{wang10}. \\
First we present our largest run: a (50 Mpc)$^3$ volume simulated with  
$2400^3$ cells and DM particles (resolution $20.8 ~ \rm kpc$), which, as far as we know, is the largest MHD cosmological simulation to date.
The simulation used $\sim 4.5$ million core hours running on 512 nodes (2048 cores in total) on Piz Daint. 
The resolution was chosen such that at least a cell size of $\sim 20$ kpc could be achieved in order to obtain sufficient amplification in $10^{14} M_{\odot}$ halos. The simulation box is large enough in order to contain massive galaxy clusters with a concordance model $\sigma_{\rm 8}$.\\

\begin{figure*}
\begin{center}
\includegraphics[width=0.99\textwidth]{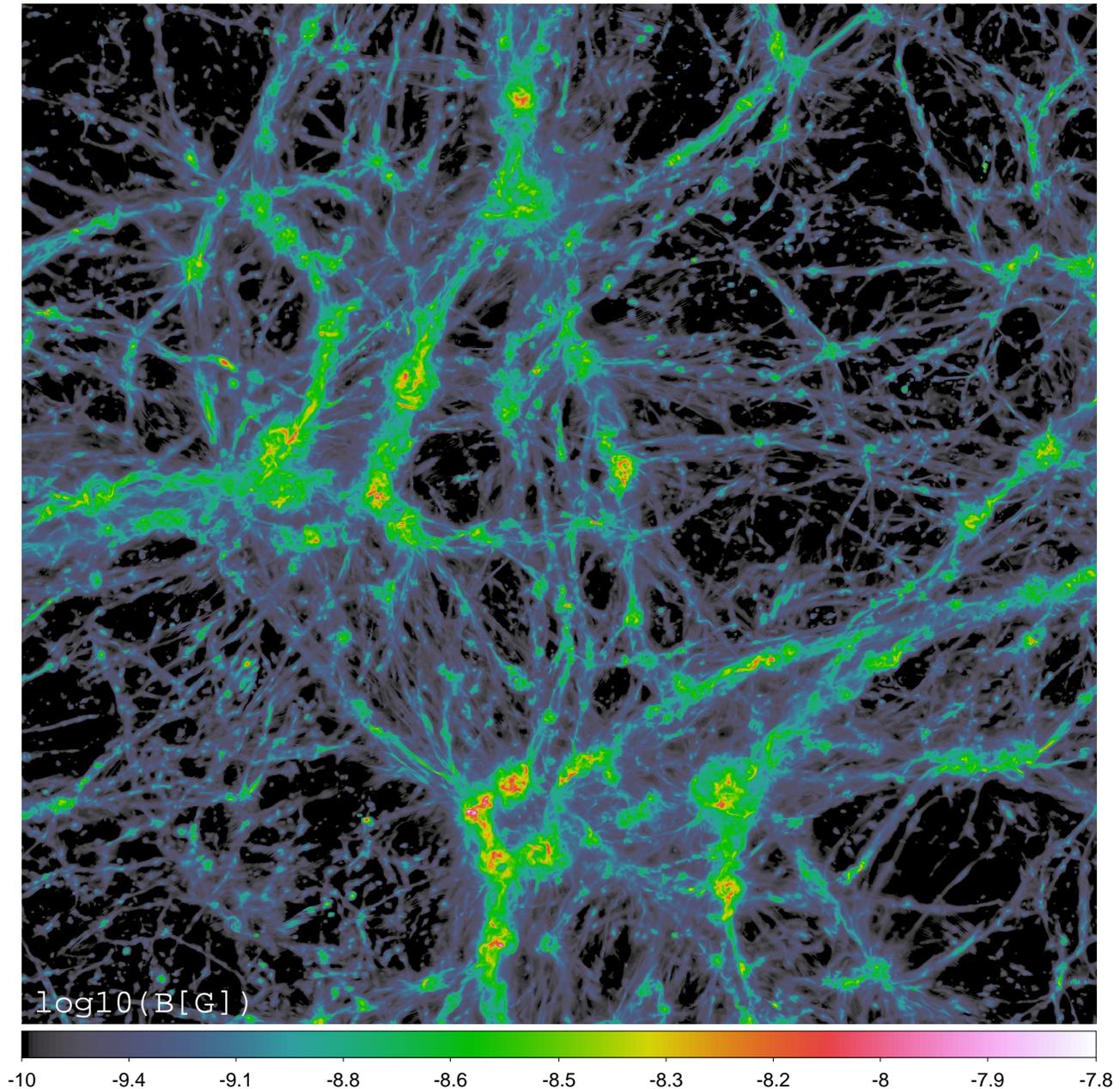}
\caption{Projected (density weighted) magnetic field intensity for our $2400^3$ simulation of a (50 Mpc)$^3$ volume at $z=0$.}
\label{fig:maps_2400}
\end{center}
\end{figure*}

Figure \ref{fig:maps_2400} shows the projected (mass-weighted) magnetic field strength at $z=0$ across the whole volume.  In regions of large over-densities, the magnetic field is amplified beyond the effect of compression by twisting motions driven by accretion and mergers. Twisted magnetic field structures are found only close to the centre of halos or in the proximity of the main
axis of filaments. The maximum field attained in filaments hardly reaches $\sim 10$ nG, while in the most massive halos the maximum magnetic field
is of the order of $\sim 0.05-0.1  ~\mu$G at most.\\

\begin{figure}
\begin{center}
\includegraphics[width=0.49\textwidth]{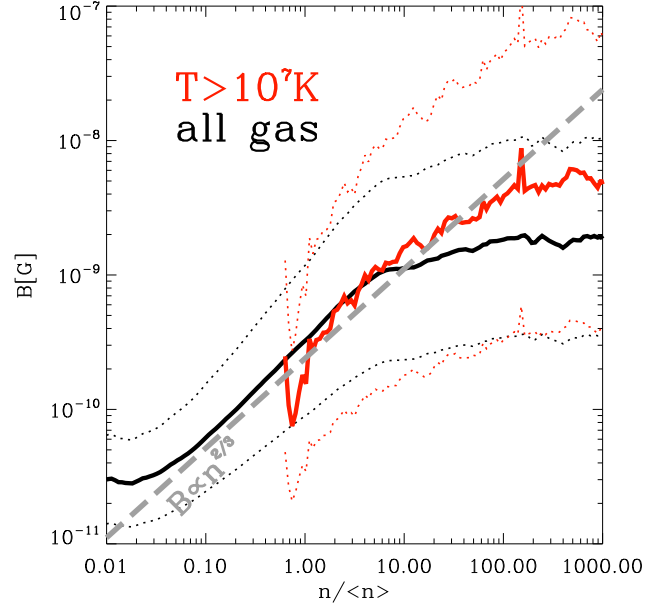}
\caption{Average relation between magnetic field and gas density for (50 Mpc)$^3$ with $2400^3$ cells at $z=0$. The black lines show
the magnetic-energy weighted average $B$ for all cells as a function of density (solid=mean, dotted=$\pm 3 \sigma$ scatter), while the red lines show the relation computed only for $T \geq 10^7 \rm K$ cells to better mark the trend in galaxy clusters. The additional grey line shows the expected trend for pure compression ($B \propto n^{2/3}$).}
\label{fig:phase_2400}
\end{center}
\end{figure}

The average $B(n)$ (Fig.~\ref{fig:phase_2400}) shows that, for the largest part, the magnetic field scales as $B \propto n^{2/3}$ with little scatter.  
At high densities, the large number of small halos dominates the average and cause a flattening of the relation because of the small number of massive hot clusters \footnote{A flat relation, $B(n)$, at high densities has also been reported by \citet{sk13} using {\enzo} MHD simulations with Constrained Transport \citep{co11}.  A possible explanation for the level of flattening measured in their and our runs is an excess of numerical magnetic reconnection due to finite spatial resolution.}. For this reason we also plot (red lines) the average relation obtained only using cells with $T \geq 10^7 \rm K$, which highlights
the ICM. Then the upper envelope of the average reaches $\sim 0.1  ~\mu G$ at densities typical of cluster centres, which is an effect of the small-scale dynamo (Sec.~\ref{subsec:cluster}).  However, for the concordance value of $\sigma_{\rm 8}=0.8$ the high-mass clusters form late in time compared to the cluster previously simulated and hence the amplification by $z=0$ is less efficient by the end of the run. The rather small final mass/size of the clusters formed in the $(50 ~ \rm Mpc)^3$ is too small to probe
large Reynolds numbers for most of the simulated objects.  Our high temperature threshold selects objects with virial masses above $\sim 5 \cdot 10^{13} -10^{14} M_{\odot}$, i.e. with virial radii around $\sim 1$ Mpc. In this case the virial radius is sampled with at least $\geq 150^3$ cells at $z=0$, and the numerical Reynolds number of the flow is  $R_{\rm e} \sim 500$ (Eq.~\ref{eq:reynolds}). This fulfils the criterion proposed by \citet{2011ApJ...731...62F} and  \citet{2013MNRAS.432..668L}, according to which a minimum amount of $128^3$ cells per Jeans length is necessary to obtain dynamo effects in primordial halos. At the over-density typical of filaments, $n/\langle n\rangle \sim 1-10$, the average magnetic field is $\leq 10$ nG, as found in our previous filament runs.

\subsection{Additional magnetic field seeding by galaxies}

Finally, we investigate the possible role of additional magnetic field seeding from galaxies crossing the filament. In a  simulation box of (25 Mpc)$^3$ sampled by a $1200^3$ mesh, we tested the effect of releasing
additional magnetic fields as small magnetic loops injected at the estimated location of forming galaxies. 
The location of each presumed galaxy was assigned based on a (comoving) gas over density larger than $500$ times the critical gas density, and at the centre of
each over-dense region we injected a magnetic loop ($3^2$ cells across) with a total magnetic field strength corresponding to $\beta=100$
at the location of each galaxy. For the sake of simplicity, we enabled the seeding from galaxies only once at $z=2$, and compared the results at $z=0$ to the model with purely primordial seeding. This model can only test the efficiency of magnetisation of filaments and galaxy clusters through stripping and mixing of gas from magnetised halos in the course of their motions inside large-scale structures. Note that our seeding model does not include the additional effect of gas outflows driven by winds and AGN.\\

\begin{figure}
\begin{center}

\includegraphics[width=0.45\textwidth]{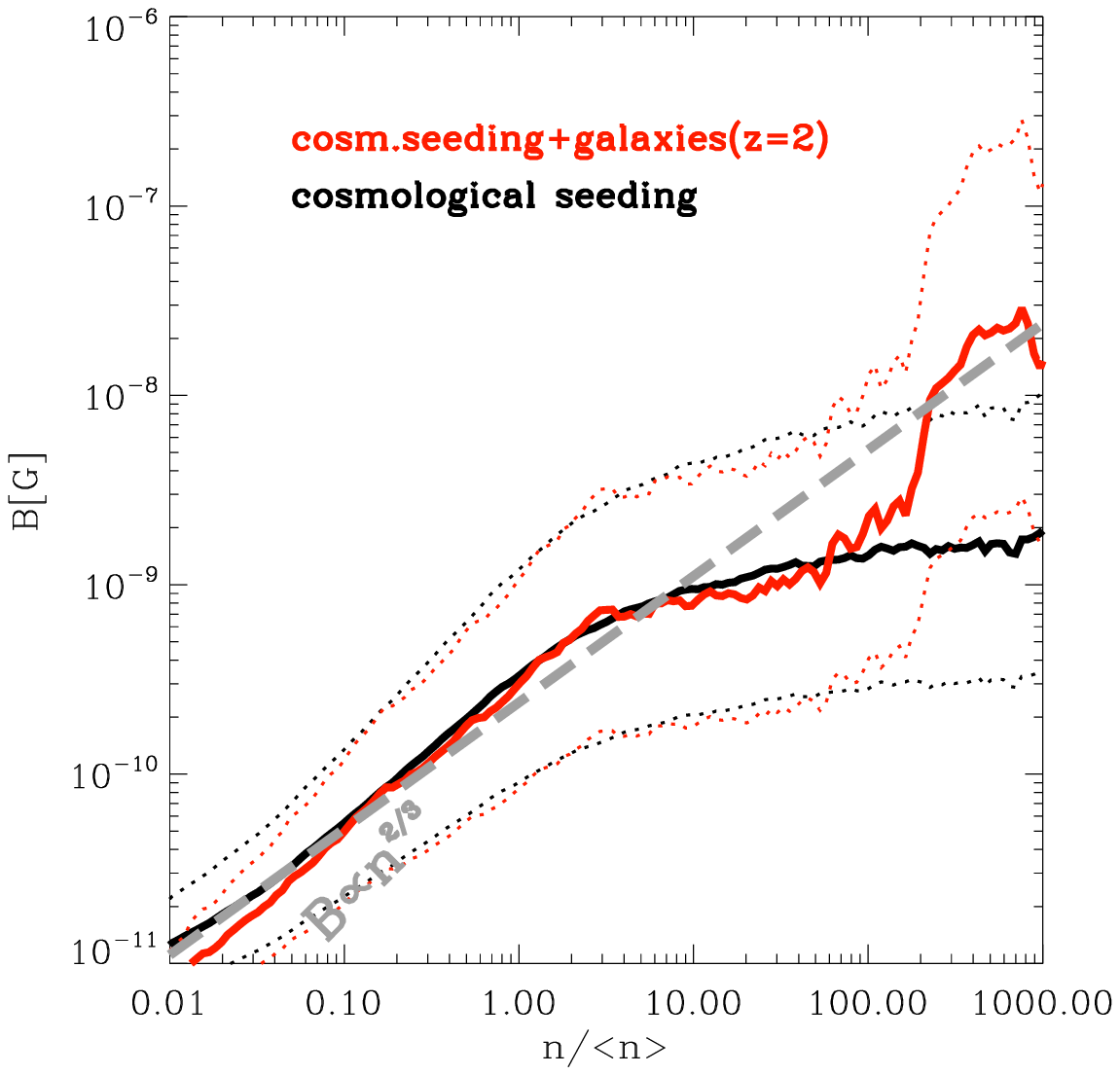}
\includegraphics[width=0.45\textwidth]{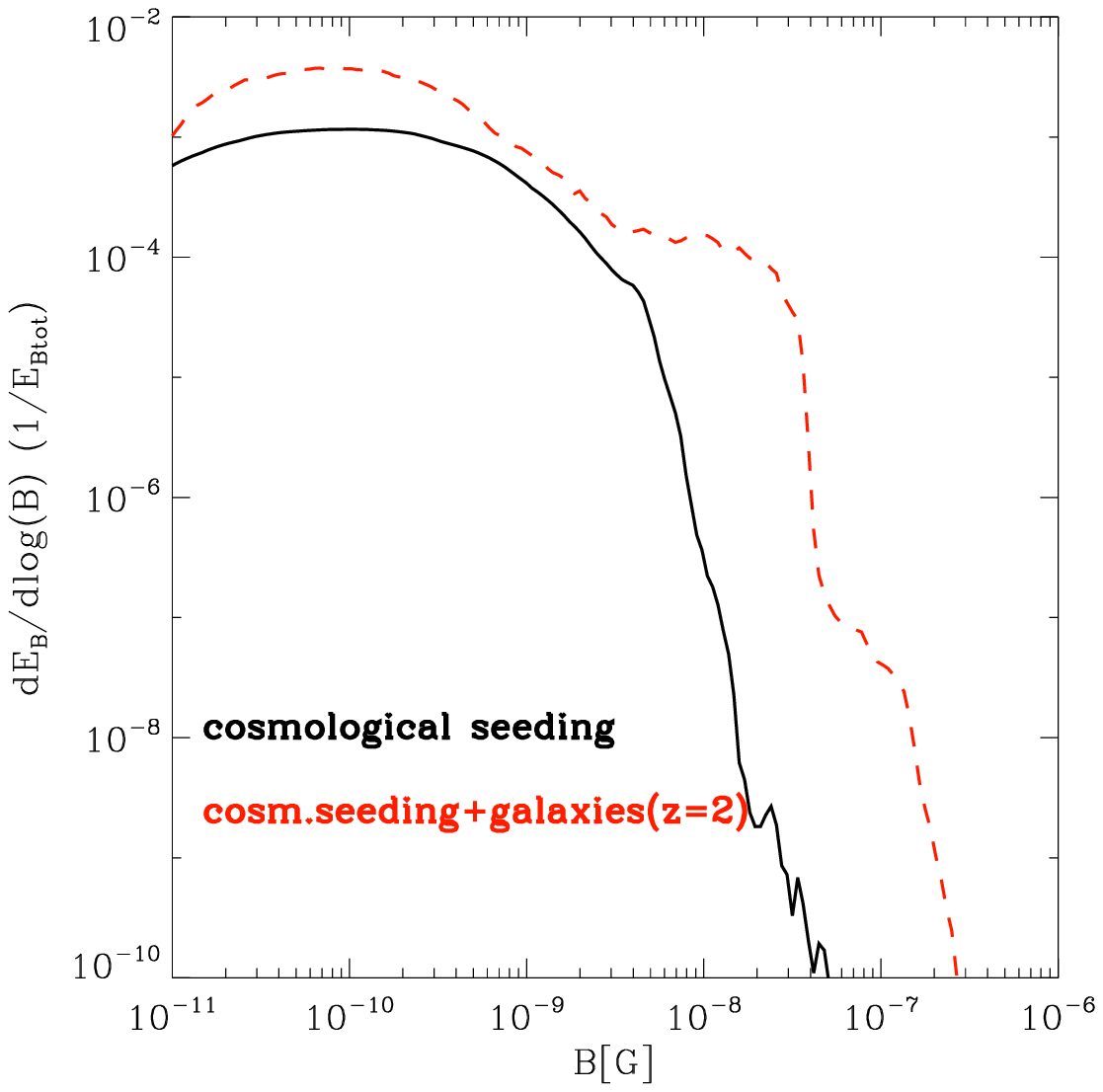}
\caption{Top panel: average relation between magnetic field and gas density (as in Fig.~\ref{fig:phase_2400}) for two resimulations of (25  Mpc)$^3$ volume at $z=0$, with a cosmological weak magnetic field initialised at $z=30$ (black) or with the additional
release of magnetic loops from "galaxies" in the volume at $z=2$ (red). The additional grey line shows the expected results for pure compression. Bottom panel: energy-weighted distribution of magnetic fields for the same runs.} 
\label{fig:distrib_1200}
\end{center}
\end{figure}

The additional seeding magnetises the high-density ICM leading to field strengths of up to
of $\sim 0.1-1 ~\rm \mu$G at the centre of the most massive halos at $z=0$. 
The distribution function of magnetic energy and the average $B(n)$  (Fig.\ref{fig:distrib_1200}) shows how the galactic seeding has the greatest effect in halos ($n/\langle n \rangle \geq 100$). There, the final magnetic field is $\sim 10-30$ times larger, reaching $\sim 0.3 \mu$G even
in the low-mass halos formed in this smaller box.  However, the effect outside of these halos is quite limited since only the $B \geq 10$ nG is significantly affected by the additional seeding from galaxies (bottom panel). 

In summary, while more complex time-dependent model of magnetic seeding from high-redshift galaxies are required, our results do not show significant large-scale magnetisation by the simple advection and stripping of magnetised galaxies. The inclusion of fast (or continuous) magnetised outflows driven by galactic
activity might yield different results. SPH simulations by \citet{donn09} have  shown that the magnetisation 
of the cosmic web outside of halos in galactic seeding scenarios is very model-dependent. 

\section{Discussion}
\label{sec:discussion}

We have investigated the amplification of primordial magnetic fields as a function of spatial resolution. Our results can be summarised as follows:

\begin{itemize}
\item {\it Magnetic field amplification in the ICM:} We have simulated the small-scale dynamo in a galaxy cluster with uniform grids of increasing resolution (from $220$ to $13 ~\rm kpc$). At resolutions with cell sizes below $\sim 26 $ kpc we observe the emergence of small-scale power in the magnetic energy spectra. The amplification seems to have reached
convergence at  the maximum resolution of  13 kpc (i.e. $\sim 1/100$ of the cluster virial radius at $z=0$), at least inside the virial region. The magnetic fields reach $\sim 0.4 ~ \rm \mu$G in the cluster core, corresponding to $\sim 1/100$ of the thermal energy of the cluster within the same volume.
Although our setup is rather artificial (due to the use of an artificially large value of $\sigma_{\rm 8}$ in order to enable the growth of a massive cluster
inside a small cosmic volume), the results are in agreement with previous results \citep[][]{do99,br05,2008A&A...482L..13D,donn09,co11}. 

\item{\it Magnetic field amplification in cosmic filaments:} In filaments, the maximum amplification factor for the magnetic energy is of the order of $\sim 100$ and the maximum field
strength, close to the axis of the filament, hardly reaches $\sim 0.01 ~\mu G$.  
The corresponding magnetic energy is only  $\sim 10^{-5}$ of the gas kinetic energy, smaller than what is found in driven turbulence simulations \citep[e.g.][]{2011PhRvL.107k4504F}. The physical reason for this is discussed in the next Section.

These results seem to be independent of resolution and apply up to the largest
Reynolds number we could probe here, $R_{\rm e} \approx 200$. The independence of resolution stems from the fact that the ratio of kinetic energy of compressive and solenoidal modes within the filament does not change significantly with resolution. Compressive forcing only leads to inefficient magnetic field amplification. 

\item{\it Amplification as a function of environment:} Inside halos where the virial volume is sampled with enough resolution elements ($\geq 150^3$ inside the virial volume) we find some dynamo amplification, as suggested by \citet{2011ApJ...731...62F} and \citet{2013MNRAS.432..668L}. The additional release of stronger magnetic fields from the high density peaks of halos (here assumed to take place only once at $z=2$) does not affect the magnetic fields in filaments at $z=0$. However, it does increase the magnetisation of the ICM at $z=0$, due to stripping and further mixing of the additional magnetic field in the turbulent ICM. 

\end{itemize}

\subsection{What is the difference between the small-scale dynamo in clusters and filaments?}

\begin{figure*}
\begin{center}
\includegraphics[width=0.95\textwidth]{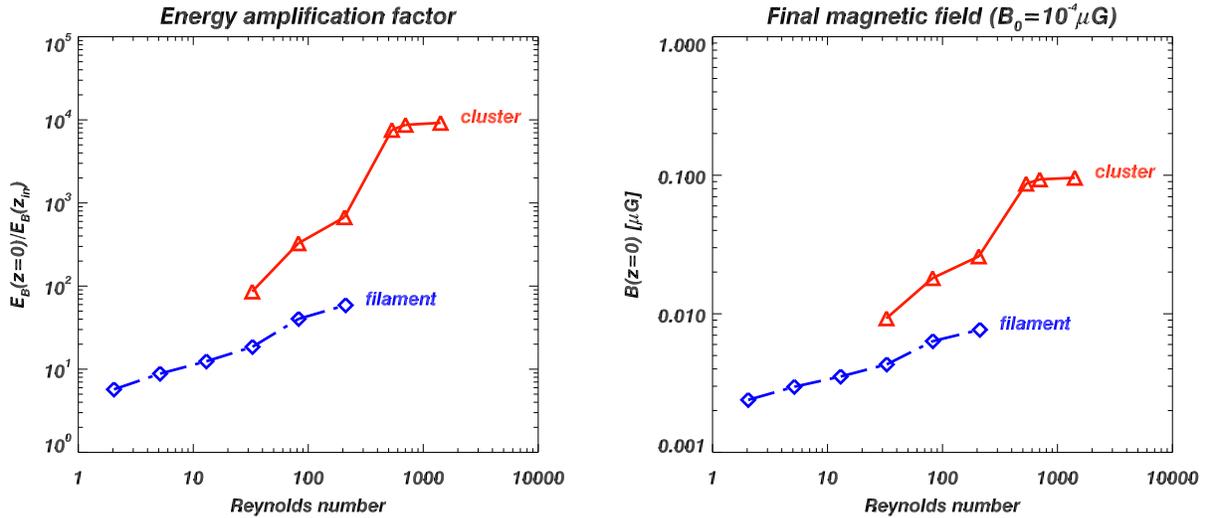}
\caption{Amplification of magnetic energy as a function of numerical resolution (top) and Reynolds numbers (centre), for the clusters and filaments. Bottom panel: average magnetic field at $z=0$, considering a uniform seed field of $B=10^{-10} \rm G$. The Reynolds numbers of each run
is computed as in Eq.~\ref{eq:reynolds}, based on the typical size of the cluster ($\sim 3$  Mpc) and
of the filament ($\sim 4 \rm Mpc$).}
\label{fig:amplification}
\end{center}
\end{figure*}

Figure \ref{fig:amplification} summarises our results for the amplification of magnetic field in the ICM and in the WHIM, as measured in our cluster and filament runs. The plots show the 
amplification of magnetic energy and of the mean magnetic field strength (averaged inside the cluster and filament volume) at $z=0$, where we assigned
a fiducial maximum Reynolds number to both systems from Eq.~\ref{eq:reynolds}. The Reynolds numbers in the filament are smaller but the observed dependence on resolution suggest that there would be no efficient dynamo, even for fairly large numerical Reynolds numbers ($\sim 200$). 
The magnetic fields in the ICM can be understood from simulations \citep[][]{2004ApJ...612..276S,2009ApJ...693.1449C,2011arXiv1108.1369J} with $P_{\rm m}=1$ (where $P_{\rm m}=\eta/\nu$ is the Prandtl number, and $\eta$, $\nu$ are the magnetic resistivity and physical viscosity, respectively). They concluded that for a large enough Reynolds number
an exponential growth of the field is observed, followed by a linear growth on the timescales of several tens of dynamical times. 
During the exponential phase $B(t) = B_0 \exp(\Gamma t/\tau)$, where $B_0$ is the initial field strength, $t$ is the 
time and $\tau$ is a characteristic time of the system (which can be here approximated as the sound crossing time). A fast dynamo occurs only when $\Gamma$ is $\gg 1$. In a $P_{\rm m}=1$ regime the relation between $\Gamma$ and the Reynolds number is $ \Gamma \approx \frac{R_{\rm e}^{1/2}}{X}$ ($X$ is a numerical factor of order $X \sim 15-30$, from which it follows that $R_{\rm e} \geq 15^2-30^2$ to enter the exponential phase). These results suggests that even if the system is subject to continuous turbulent forcing at the largest scales, it takes several tens of crossing time for the system to reach a stationary magnetic field strength of the order of 
 $\sim 30$ percent of the total kinetic energy. This is not far from what we observe, at least on the smallest spatial scales, in our simulated cluster at the highest resolution, owing to the fairly large $R_{\rm e} \sim 1400$ there. This is not observed in the filament, even at our highest resolution, with no sign of dynamo action.\\

Besides the smaller numerical Reynolds number in the filaments, there are additional reasons to believe that the amplification cannot be significantly
larger than this - even in case of a much larger $R_{\rm e}$. First, previous simulations by independent groups have shown that compressive forcing of turbulence is very inefficient in producing dynamo amplification, as most of the energy pumped into the system is quickly dissipated into shocks \citep[][]{2006PhFl...18g5106H,2011PhRvL.107k4504F,2011arXiv1108.1369J} (see also our Appendix). In particular, \citet{2011PhRvL.107k4504F} have shown that the magnetic field dynamo driven by forced turbulence in a box exhibits a characteristic drop of the growth rate at the transition from subsonic to supersonic turbulent flow. Solenoidal turbulence drives more efficient dynamos, due to the higher level of vorticity generation and the stronger tangling of the magnetic field. Based on the different approach of solving the Kazantsev equation with the WKB (Wentzel, Kramers, and Brillouin) approximation, \citet{2012PhRvE..85b6303S} measured the growth rate of magnetic field dynamo in different turbulent models. They showed that for highly compressible turbulence the critical Reynolds number to produce an efficient dynamo is larger than in the case of Kolmogorov turbulence (i.e. $\sim 2700$ vs $\sim 100$), and that the growth rate in the compressible case has a shallower dependence on the Reynolds number (i.e. $\Gamma \propto R_{\rm e}^{1/3}$ for Burgers turbulence and $\propto R_{\rm e}^{1/2}$ for Kolmogorov turbulence
). Finally, using a Fokker-Planck approach to compute the growth of magnetic field dynamo in the non-linear regime, \citet{2013NJPh...15b3017S}  have recently shown that the characteristic length scale of the magnetic field grows faster in Burgers than in Kolmogorov turbulence. This confirms that in the presence of compressive forcing dynamo amplification is much less efficient than in the solenoidal forcing case. This is even more apparent in filaments because strong advection motions along the spine of filaments continuously move turbulent eddies away from the region of colliding flows, reducing small-scale dynamo more severely than in the case of stationary forcing of solenoidal turbulence in the box \citep[e.g.][]{2011PhRvL.107k4504F}.
Here, we have analysed how the modes of the velocity field evolve with
spatial resolution in both cases. To this end, we decomposed in the velocity field using the Hodge-Helmholtz projection in Fourier space \citep[e.g.][]{kr11}, and computed the kinetic energy in the compressive and in the solenoidal modes. In both cases, we selected a region at $z=0$, not affected by infall motions outside of accretion shocks. \\
The panels in Figure \ref{fig:fila_modes} show our results: while in the ICM  
the budget of kinetic energy in compressive modes decreases with resolution, in the WHIM the energy does not change. At the best resolution, the energy in compressive modes is only $\sim 30$ percent in the ICM and up to $\sim 60$ percent in the WHIM. The fact that the kinetic energy in solenoidal motions is higher 
in galaxy clusters and smaller in the WHIM has already been established by cosmological numerical simulations \citep[][]{ry08,2011MNRAS.414.2297I,2013ApJ...777...48Z,2014ApJ...782...21M}.
However, we find that this persists at high resolution and that
in filaments $\sim 2/3$ of the kinetic energy is in form of supersonic compressive modes. 
This explains the lack of amplification in filaments, despite the increase in the numerical Reynolds number. Indeed,
the maximum amplification of magnetic energy in subsonic solenoidal turbulence (as in the ICM) is expected to be $\sim 1-2$ orders of magnitude higher than the maximum amplification reached in supersonic compressive turbulence (as in the WHIM) \citep{2011PhRvL.107k4504F}. Moreover, the growth 
rate in the first case is $\sim 5-10$ times faster.

\subsection{Physical and numerical limitations of the MHD picture}

Our resolution tests went down to a resolution of $\sim 20$ kpc even though the smallest collisional scales in the WHIM should be $\sim 100 - 10^3$ kpc based on pure Coulomb interactions. Below this
scale a kinetic modelling could be more appropriate \citep[][]{2013arXiv1304.3941W}. However, if efficient scattering occurs between particles and magnetic perturbations induced by small scale plasma instabilities, then the mean free path of particles decreases in a 
self-regulating process: if turbulence is stronger at the scale of injection,
the mean free path of plasma particles is reduced and the range
of scales over which the fluid behaves as collisional is increased \citep[][]{sch05,2011MNRAS.410.2446K,bl11}. Whether or not the same
picture applies to the even more tenuous and weakly magnetised WHIM in filaments, is presently uncertain. \\

Regarding the MHD scheme, the Dedner hyperbolic cleaning scheme \citep[][]{ded02} is a robust and widely used method in the literature, but is prone to small-scale artefacts and artificial dissipation due to the $\nabla \cdot \bar v$ wave necessary to 
limit to the presence of magnetic monopoles. In the literature, this method has been compared to others, both for grid and SPH simulations \citep[][]{ded02,wa09,2010JCoPh.229.5896M,kr11,2013MNRAS.428...13S,2014ApJ...783L..20P}, reporting good consistency. In particular, \citet{kr11} have investigated in detail the performance of several MHD methods in the case of decaying supersonic turbulence in an isothermal box, including
the Dedner scheme implemented in {\enzo}. They concluded that all codes agreed well on the kinetic and magnetic energy decay rate, but they 
varied on the amplitude of the peak magnetic energy, as this was significantly dependent on the numerical dissipation of each method (that in turns determines the effective magnetic Reynolds number). They found that the use of explicit divergence cleaning  reduces the magnetic spectral bandwidth relative to codes that preserve the condition on the magnetic field exactly, as the constrained transport (CT) methods. They concluded that codes that fall short in some of the
investigated diagnostics (i.e. dissipation of small scale modes in the Dedner cleaning scheme) still can get to the correct physical answer, provided that
they compensate the higher numerical dissipation with higher numerical resolution.

\subsection{Comparison to previous work}

Our results for non-radiative runs seem to be in agreement with those obtained by \citet{br05}, \citet{2008A&A...482L..13D} and \citet{co11}, who also reported evidence of growth of magnetic fields in excess of simple compression even if with lower efficiencies \footnote{Also the latest very high-resolution cluster simulations by F. Miniati \citep[see][for a study of the hydrodynamical properties of these simulations]{2014ApJ...782...21M} confirms the difficulty to get to very large amplification factors, despite the fairly large numerical 
Reynolds number achieved there (private communication).}. Runs with radiative cooling readily obtain
 magnetic fields of the order of $\sim \mu G$ in the ICM, mainly as a result of overcooling \citep[][]{2008A&A...482L..13D,co11,ruszkowski11}. Despite some similarity
in the magnetic spectra, it is difficult to relate to the {\enzo}-MHD simulations by \citep[][]{xu09,2011ApJ...739...77X} since their seeding is very different from ours.\\

There is disagreement, though, with the results of cosmological SPH
simulations \citep[][]{do99,2007MNRAS.375..657G,do08,donn09,2009MNRAS.398.1678D,bo11,beck12,2013MNRAS.428...13S,beck13}, that typically reach much larger amplification factors for the magnetic energy, already at high redshift ($z \geq 2$). Understanding these differences is beyond the goal of this paper, and we can only speculate that the reason
lies in the capability of SPH in refining the innermost regions of halos already at earlier times. 
However, also the difficulty in correctly modelling small-scale velocity structures (and the connected magnetic-field amplification) in SPH might
 be responsible for the difference \citep[][]{2012MNRAS.423.2558B,2012MNRAS.420L..33P}, for which ad-hoc solution are required \citep[][]{do05,2009MNRAS.398.1678D,2013MNRAS.429.3564D,2013MNRAS.428...13S}.\\

Few papers address the magnetic field amplification in filaments.
Early MHD grid simulations by \citet{Sigl:2003ay} predicted $\sim 10-100 ~\rm nG$ fields in filaments. However, the total normalisation of the magnetic fields
had to be scaled up in order to match the observation of the Coma cluster. Taking this into account and normalising by the assumed initial seed field, these simulations essentially showed only compressive amplification of magnetic fields in filaments, in line with what we also find at low resolution.
\citet{br05} applied instead AMR and a passive scheme in {\small FLASH} to monitor the amplification of magnetic fields also at the scale of filaments, and
reported an average amplification factor of $\sim 10^3-10^4$ for the magnetic energy of filaments, i.e. larger than what we found here. This can be explained by the difference in the adopted MHD scheme, even if the spectra of magnetic fields did not show evidence for small-scale dynamo amplification and the topology of magnetic fields in filaments was found to be laminar.\\
Smaller amplification factors for the magnetic energy, essentially in agreement with our results here, were found using SPH simulations by
\citet[][]{2004JETPL..79..583D} with a constrained realisation of the nearby (100 Mpc)$^3$ Universe. 
Finally, several of our results have already been explained by \citet{ry08}, who used an hybrid approach to rescale the magnetic field distribution
obtained with a passive MHD solver coupled to a cosmological simulations. In post-processing, they then estimated  the saturated growth of magnetic fields based on the (unresolved) turbulent decay of vortical motions resolved in the simulation. The important difference in the modes of turbulent forcing in filaments
and galaxy clusters, and its impact on the amplification of weak primordial fields was already pointed out in their work, and our direct simulation with a larger spatial resolution confirmed their main results \citet{ry08}.
However, our simulations (see also our tests in the Appendix) have shown that the results of dynamo amplification in driven turbulence (specially in the isothermal case) cannot be trusted to exactly predict the maximum dynamo amplification in WHIM. First, due to the major role played by shocks even in the filament interiors, which cannot be fully captured with isothermal computations, since this largely underestimates the role of the baroclinic generation of vorticity. And, second, because of the presence of strong longitudinal motions along the filament that prevents the continuous build-up of small scale magnetic field at any specify location within the filament, as instead observed at the centre of clusters.

\begin{figure}
\begin{center}
\includegraphics[width=0.49\textwidth]{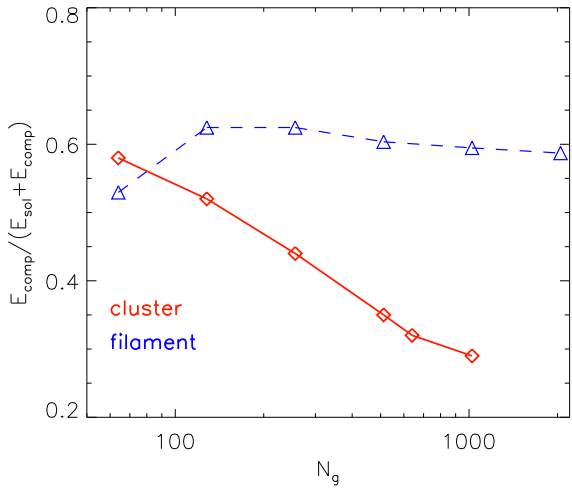}
\caption{Resolution-dependence of the ratio between compressive and total (compressive+solenoidal) kinetic energy in clusters and filaments.}
\label{fig:fila_modes}
\end{center}
\end{figure}

\section{Conclusions}

We have studied the amplification of primordial magnetic fields via a small-scale turbulent dynamo using direct MHD numerical simulations with {\enzo} \citep[][]{wa09,wang10,enzo13}. In particular, we have investigated the amplification of magnetic fields in the ICM and in the WHIM of filaments. 
While in the ICM we confirm that turbulence from structure formation can produce significant dynamo amplification (even if the measured efficiency is smaller than what is reported in some papers), in filaments we do not observe significant dynamo amplification,
even though we reached Reynolds numbers of $R_{\rm e} \sim 200$. The maximum amplification for large filaments is  of the order of $\sim 100$ for the magnetic energy, mostly due to strong compression in supersonic flows,  corresponding to a typical field of a few $\sim$ nG.  This result is independent of
resolution and follows from the inefficiency of supersonic motions in the WHIM in triggering solenoidal modes, while compressive modes are dominant in filaments at all investigated resolutions.\\

Our results can serve as a guideline for the minimum resolution for the onset of small-scale dynamo in cosmological simulations. Our results for the ICM (Sec.\ref{subsec:cluster}) suggest that a dynamical range of 
at least $L/\Delta x\sim 210$ (where $L$ is the scale for the driving of turbulence and $\Delta x$ is the numerical resolution) is necessary to observe the build-up of small-scale magnetic field in a dynamo process, as this would enable a flow with $R_e \geq 500$. Even if the bulk of turbulence injection in the ICM at late
redshift happens through mergers and on scales of a fraction of the virial radius \citep[][]{va09turbo,va11turbo,va12filter}, the converging accretion flows within and the injection of vorticity at accretion shocks \citep[][]{ry08,2014ApJ...782...21M} are likely to build up magnetic fields in the ICM on scales up to the order of the virial radius. Assuming $L \approx 2 R_{\rm v}$, the above criterion suggests that $\Delta x/R_{\rm v} \leq 100$ to have efficient dynamo, i.e. in order 
to achieve a large enough Reynolds number for small-scale dynamo, a cosmological simulation needs a spatial resolution of order $\sim 30 ~ \rm kpc$ for a $10^{15} \rm M_{\odot}$ halo ($R_{\rm v} \approx 3 ~\rm Mpc$),  of 
order $\sim 10 ~ \rm kpc$ for a $10^{14} \rm M_{\odot}$ halo, and of order $\sim 3 ~ \rm kpc$ for  $10^{13} \rm M_{\odot}$. The fact that clusters form late, combined with the fact that mergers typical inject energy at scales below $R_{\rm v}$, likely makes the above estimate a lower limit on the required resolution, as shown in our larger cosmological run (Sec.\ref{subsec:cosmo}).

It is more difficult to draw firm conclusions in the case of filaments, as our runs do not clearly show a convergence on the dynamo process. Our results suggest that a dynamical resolution equal or larger than $L/\Delta x\sim 60$ (where $L$ is the width of the filament) is necessary to approach convergence in the energy share of solenoidal and compressive motions (Fig.\ref{fig:fila_modes}), which sets the nature of the turbulent forcing within the system. Despite the fact that the theoretical Reynolds number available to the flow is large, $R_e \geq 10^2$, in the presence of this dominant compressive forcing no clear evidence of a fast dynamo is detected, even in our highest resolution runs, where the total magnetic energy is only $\sim 10^{-5}$ of the kinetic energy and carries memory of the initial magnetic field imposed at $z=30$.

The observational consequences of these results are important.
First, the deflection of UHECRs by filaments in the cosmic web is expected to be fairly small, i.e. $\leq 1$ degree, allowing the identification of extragalactic sources \citep[][]{2003PhRvD..68d3002S,2010ApJ...710.1422R}. Secondly, the detection of synchrotron emission by
electrons accelerated by shocks surrounding filaments will be very challenging since diffusive shock acceleration requires a minimum magnetic
field of $\sim 0.05-0.1 ~\mu$ G (Vazza et al., submitted). 
Within the present uncertainties about the magnetisation level of the WHIM, we suggest that {\it any} observation of large-scale fields in filaments in the radio band will 
contain valuable information about the strength of primordial magnetic fields \citep[e.g.][]{2010Sci...328...73N,wi11}. Since the growth of primordial magnetic fields in filaments should be dominated by simple compression and small-scale shocks, the dynamical memory of the system should persist over long cosmological times, and any observed magnetisation level should closely connect to the primordial magnetisation. This is different from galaxy clusters where most of the magnetic energy is extracted from the kinetic energy budget, thereby quickly erasing previous dynamical information. \\

Finally, we stress that our results imply by no means that the quest for higher resolution in the filamentary structures of the cosmic web is useless.
Provided that MHD can still be applied there (Sec.~\ref{sec:discussion}), resolution can significantly impact the Faraday Rotation from the intergalactic medium (IGM), which the SKA might probe \citep[e.g.][]{2014arXiv1406.3871A}. It also affects the
synchrotron emission from the cosmic web \citep[][]{2011JApA...32..577B,2012MNRAS.423.2325A} because shock statistics change with resolution \citep[][]{va11comparison}. The use of high resolution also allows to model galaxy formation processes in filamentary environments in detail, which is
crucial to study the impact of magnetised outflows from galaxies \citep[][]{xu09,donn09,beck13}. 

\section{acknowledgements}

Computations described in this work were performed using the {\enzo} code (http://enzo-project.org), which is the product of a collaborative effort of scientists at many universities and national laboratories. We gratefully acknowledge the {\enzo} development group for providing extremely helpful and well-maintained on-line documentation and tutorials.\\
Most of the simulations of these work have been ran on Piz-Daint at CSCS-ETH, under the CHRONOS program 2014, which we gratefully acknowledge. 
F.V. and M. B. also acknowledge the  usage of computational resources on the JUROPA cluster at the at the Juelich Supercomputing Centre (JSC), under project no. 5018 and 7006.
F.V. and M.B. acknowledge support from the grant FOR1254 from the Deutsche Forschungsgemeinschaft. 
We thankfully acknowledge J. Donnert, D. Schleicher, M. A. Latif, K. Subramanian and F. Miniati for very fruitful scientific discussion. We thank L. Vazza for his late but fundamental reading of the manuscript.

\bibliographystyle{mnras}
\bibliography{franco}

\appendix

\section{Tests of turbulence in a box }

In this appendix we show results from simulations of dynamo amplification in driven turbulence with {\enzo}. 
A thorough analysis of driven turbulence experiments has been given by \citep[][]{2004ApJ...612..276S,ry08,2009ApJ...693.1449C,2011arXiv1108.1369J,2011PhRvL.107k4504F}. Our runs serve as a proof of concept for small-scale dynamo in flows with large numerical Reynolds number, and as a preliminary benchmark for the application
of the {\enzo}-MHD algorithm to larger cosmological simulations.\\

\begin{figure}
\begin{center}
\includegraphics[width=0.45\textwidth]{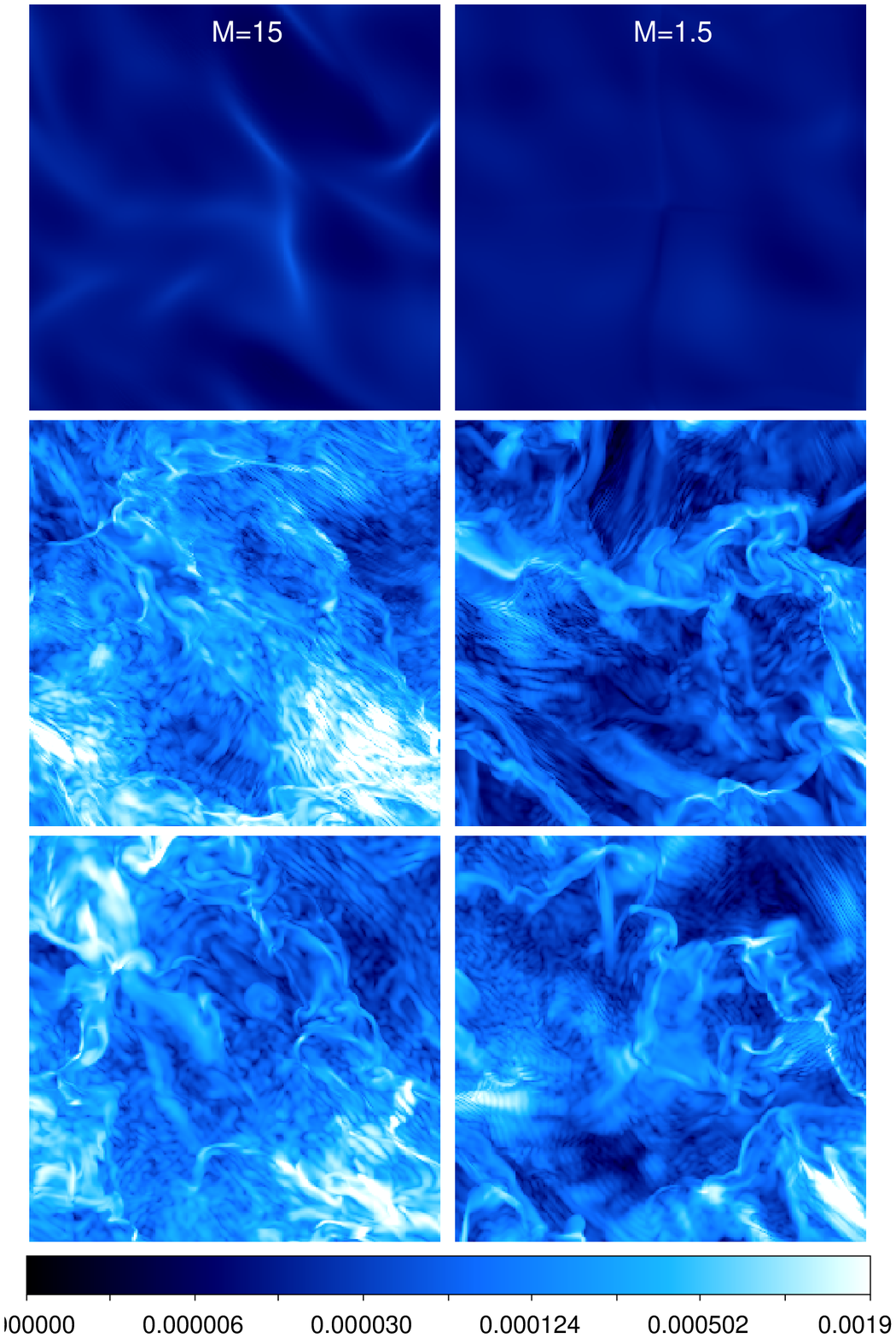}
\caption{Maps of magnetic field strength (arbitrary units) for a central  slice in our driven turbulence tests with $512^3$, for the $M=15$ forcing (left) and the $M=1.5$ forcing (right) case, at the epochs of  $\approx 0.005 t_{\rm dyn}$, $\approx 1.5 t_{\rm dyn}$ and $\approx 3 t_{\rm dyn}$, where $t_{\rm dyn}$ is the dynamical time.}
\label{fig:box1}
\end{center}
\end{figure}

\begin{figure}
\begin{center}
\includegraphics[width=0.495\textwidth]{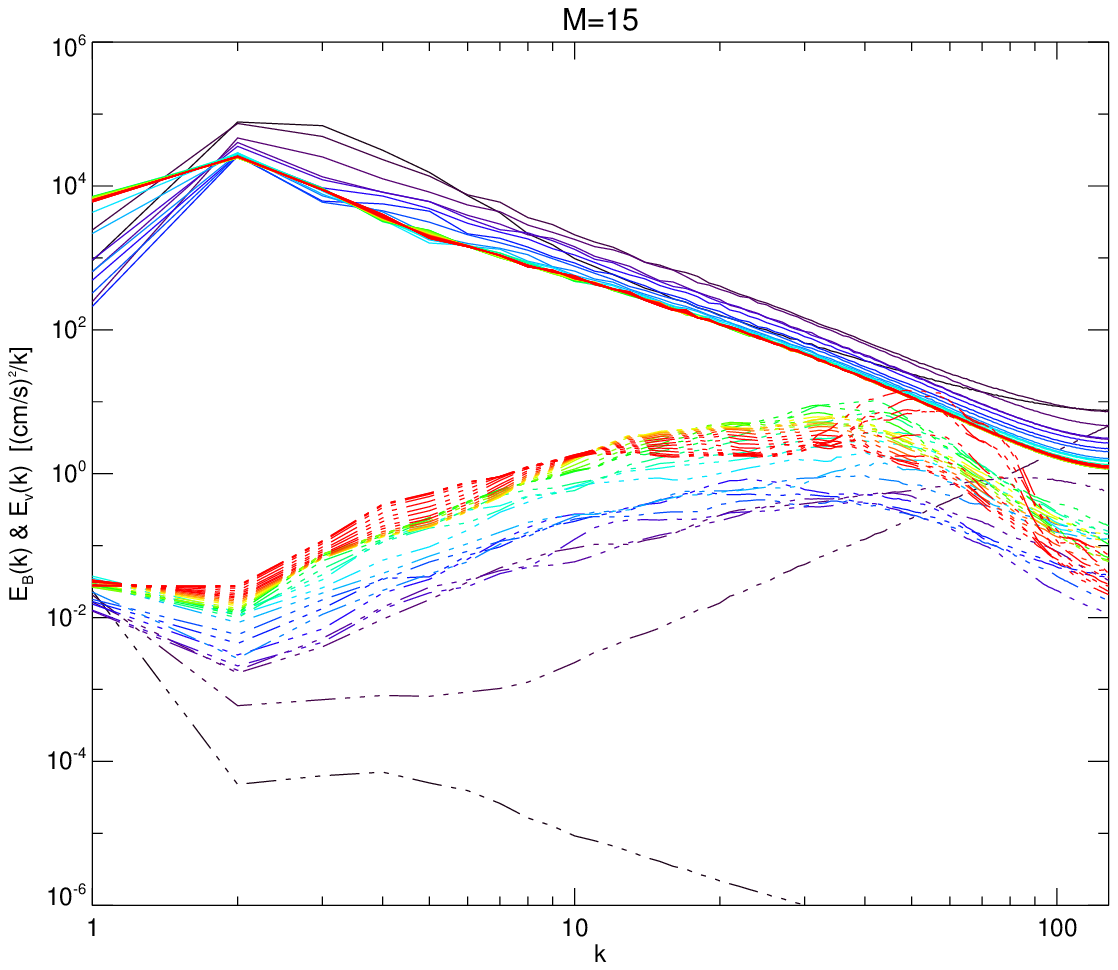}
\includegraphics[width=0.495\textwidth]{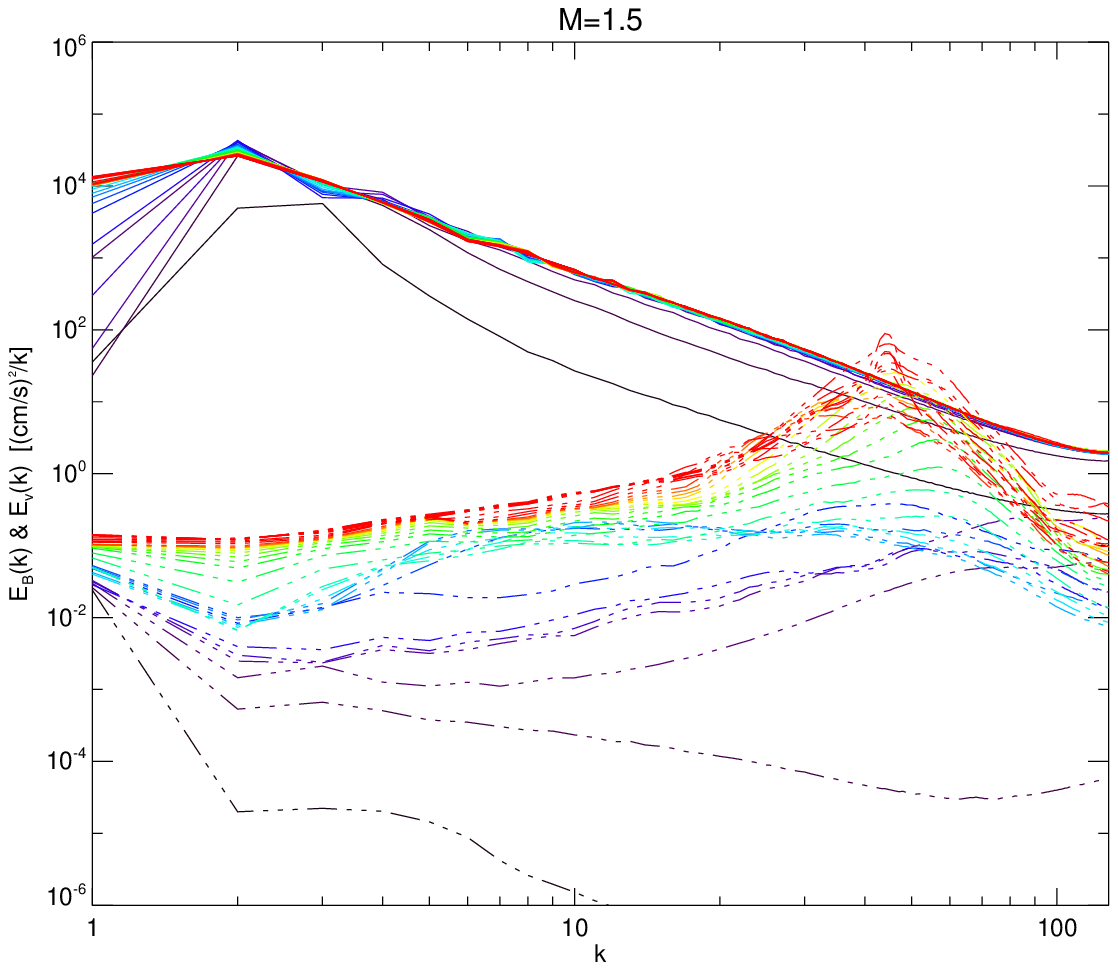}
\caption{Velocity power spectra (solid lines) and magnetic power spectra (dot-dashed lines) for two $512^3$ ``turbulence in a box'' runs, assuming a constant forcing
of $M=15$ (left) and of $M=1.5$ (right). All spectra are computed within a $256^3$ sub volume contained in the two boxes. The time evolution samples 
$\sim 4 t_{\rm dyn}$ with roughly constant time spacing.}
\label{fig:box}
\end{center}
\end{figure}

In detail, we simulated the evolution of MHD turbulence in a regular box starting with uniform density, temperature and passive uniform magnetic fields, equation
of state $\gamma=5/3$, and applied continuous stirring from turbulence injected at $k=2$ (where $k=1$ corresponds to the box size). This is 
done with a specific module available in {\enzo}, that generates random isotropic velocity fields with specified input spectra and absolute
normalisation for the total velocity field \citep{wang10}. \\
In our tests, we employed $512^3$ boxes and drove $M=1.5$ and $M=15$ isotropic motions in a continuous way. 
Figure \ref{fig:box1} shows the magnetic field strength at three different times for these two runs, at epochs $\approx 0.005 t_{\rm dyn}$, $\approx 1.5 t_{\rm dyn}$ and $\approx 3 t_{\rm dyn}$, where the dynamical times is defined as $t_{\rm dyn}=L_{\rm box}/V_{\rm drive}$ ($L_{\rm box}$ is the box size and $V_{\rm driv}=M c_{\rm s}$ is the rms velocity at the forcing scale. \\
The evolution of kinetic and magnetic spectra until $4 \cdot t_{\rm dun}$ for the two cases is given in Fig.~\ref{fig:box}, and highlights the significantly different evolution of magnetic field structure in the two regimes. 
In the $M=15$ case  after a very tiny fraction ($10^{-2}$) of the dynamic time we see the emergence
of magnetic energy on very small scales, as an effect of shocks that are formed very early inside the box due to strong supersonic motions. The small-scale magnetic energy increases over time, without significantly changing the location of the peak of magnetic energy, and after $\sim 4 t_{\rm dyn}$ we observe the hint of equipartition with kinetic energy on the smallest scales. This case is close to the case of the WHIM in cosmic filaments, due to the 
involved supersonic flow, even if the multiple collisions of oblique shocks are more efficient in driving solenoidal motions in the medium (mostly through baroclinic generation of vorticity and at curved shocks through Crocco's theorem, e.g. \citealt{2011arXiv1108.1369J}), which reaches roughly a $\sim 50$ percent budget of the total kinetic energy at the end of the run, i.e. much more than in our simulated filament. 
Moreover, the forcing to which the magnetic eddies are subjected is
constant in time, while in the case of filaments (Sec.~\ref{subsec:fila}) strong advection motions longitudinal to the major axis of the filament tend to
continuously replace magnetic eddies at a given Eulerian location, thereby reducing their growth rate. \\
Conversely, the $M=1.5$ is closer to the case of the simulated ICM, given the transonic forcing regime and the enhanced presence of solenoidal motions 
by the end of the run ($\sim 60$) percent of the total kinetic energy, i.e. similar to our high-resolution ICM runs. In this case we observe in the spectra a slower build-up of small scale magnetic energy, and the progressive increase of the total velocity spectrum over time. In this transonic forcing the thermalisation of kinetic energy at shocks is obviously greatly reduced, and a more volume filling and tangled velocity field can build over time. 
Roughly after one dynamical time, we observe the formation of a well defined peak in the magnetic spectrum, that progressively moves to larger
spatial scales and becomes of the same order the kinetic energy at the smallest scales, as predicted in efficient small-scale dynamo \citep[][]{2004ApJ...612..276S,2009ApJ...693.1449C}.\\
Both simulations confirm the possibility of simulating small-scale dynamo amplification with the {\enzo}-MHD version we adopted to obtain our results in the main paper, and suggest that to get to more quantitative answers in the case of the ICM and of the WHIM one must resort to proper 3D cosmological simulation, in order to have the large-scale dynamics properly taken into account.

\end{document}